\documentclass[aip,amsmath,amssymb,
 reprint,   
]{revtex4-2}
\usepackage{amssymb,amsthm,amsmath}
\usepackage{xcolor,paralist,hyperref,titlesec,fancyhdr,etoolbox}
\usepackage{graphicx}

\begin{document}


\title{Conditional Neural Field based Reduced Order Model for Dynamic Ditching Load Prediction}



\author{Henning Schwarz}
\homepage{Author to whom correspondence should be addressed. Electronic mail: henning.schwarz@tuhh.de}
\author{Pyei Phyo Lin}
\affiliation{Institute for Fluid Dynamics and Ship Theory, Hamburg University of Technology, Am Schwarzenberg-Campus 4, D-21073 Hamburg, Germany}
\author{Jens-Peter M. Zemke}
\affiliation{Institute of Mathematics, Hamburg University of Technology, Am Schwarzenberg-Campus 3, D-21073 Hamburg, Germany}
\author{Thomas Rung}
\affiliation{Institute for Fluid Dynamics and Ship Theory, Hamburg University of Technology, Am Schwarzenberg-Campus 4, D-21073 Hamburg, Germany}



\date{\today}

\begin{abstract}
    Grid-based neural networks such as convolutional autoencoders are widely used in dimension reduction-based surrogate models for computational fluid dynamics. 
In recent years, the use of coordinate-based approaches like conditional neural fields has emerged. Their independence of the spatial discretization is a beneficial feature for various applications in computational fluid dynamics. 
    This paper discusses the spatio-temporal prediction of aircraft ditching loads using a conditional neural field approach. 
    The model is evaluated using two datasets for the dynamic loads of the fuselage of a DLR-D150 aircraft, one of which relates to a single fixed spatial discretization and the other that includes data from different discretizations.
    When paired with a long short-term memory (LSTM) network in the latent space, the neural field-based model achieves a spatio-temporal prediction accuracy for the first data set that is close to that of grid-dependent convolutional autoencoder-based models, and with  significantly less parameters. 
    Results for the second data set demonstrate the ability of the neural field-based approach to reconstruct ditching loads accurately for heterogeneous spatial discretizations. This allows for flexible use of training datasets generated for different geometries and/or discretizations, as well as the use of the surrogate model to predict loads for different configurations.
\end{abstract}


\maketitle 

\section{Introduction}
Machine learning (ML) has increasingly been applied in the fluid dynamics community over the last decade with great success in a variety of areas, with reduced order modeling being one of them. Next to classical approaches like the proper orthogonal decomposition (POD) \cite{lumley1967structure}, surrogate models based on (variational) autoencoders have been studied on a wide range of applications \cite{Agostini20, eivazi:2020, Pache22, eivazi:2022, kang:2022, solera-rico:2024, schwarz:2024, loft:2025}, whereby models such as long short-term memory (LSTM) \cite{hochreiter:1997} or transformer networks \cite{vaswani:2017} were applied in the latent space to even predict temporal dynamics \cite{eivazi:2020,wu:2021,Hemmasian:2023,solera-rico:2024}.

In many previous works, convolutional autoencoders (CAEs) are used as they can efficiently capture local dependencies inherent in the physical data. 
Although CAE-based models are effective, they require the same (spatial) discretization for all data examples in the data set. 
Therefore, further data processing is required for grids that differ in the examples of the dataset.
Meta grids can be employed in this regard \cite{Pache22, loft:2025, maram2026adjointbasedshapeoptimizationship}, but involve additional effort and difficult-to-control errors. Furthermore, the possibility of flexibly using the ML model for different configurations would be appreciated in engineering practice.  Consequently, there is strong motivation to train surrogate models directly on unstructured data, i.e., heterogeneous grids. 
Several works address this issue with graph neural networks (GNNs) \cite{pfaff2021learning, brandstetter2022message, BARWEY2023112537, franco:2023}. Although these might work well in combination with different grids, there are also known issues, e.g., difficulties to generalize to meshes not seen during training \cite{serrano:2023},  which presents a major hurdle in engineering practice.

Recently, the use of neural fields for surrogate modeling has emerged. Neural fields are coordinate-based networks, i.e., 
they map the spatial (and potentially temporal) coordinates of one point to a specific output quantity at this point\cite{Xie:2022}. They are therefore independent of the discretization, which is a highly useful feature for computational fluid dynamics surrogate models.
In neural field-based surrogate models, the output is often conditioned to a provided or learned latent vector to enable considering different cases with a single model, similar to autoencoder-based models. An autodecoding approach \cite{Park_2019_CVPR} is usually employed to learn the latent vector and the model is referred to as a conditional neural field (CNF).
CNFs have been applied in several works for fluid dynamics problems. 
For example, the DINo method \cite{yin2023continuous} combines a CNF and a neural ODE for spatio-temporal predictions of solutions to partial differential equations. 
The CORAL method \cite{serrano:2023} combines two CNFs, representing input and output functions, and a latent space processor, which maps the latent input representation to the latent output representation, for operator learning on general geometries.
 The more recent CoNFiLD method employs a CNF combined with a latent diffusion model to generate spatio-temporal turbulence \cite{du:2024}. Guo et al. benchmark CNFs with different conditioning mechanisms against POD and CAEs for dimension reduction of turbulent flows \cite{guo:2026}. 
Catalani et al. use CNFs in their proposed MARIO framework for aerodynamic surrogate modeling \cite{CATALANI2026106929}. 
They use available operating and geometric conditions directly to define the latent vector. If the dataset has geometric variability without access to the corresponding true parameters, the part of the latent vector providing the geometric information is optionally learned.

Connections between CNFs and neural operators, which have achieved state of the art performance in operator learning \cite{li2021fourier}, have been reported by Wang et al. \cite{wang2025cvitcontinuousvisiontransformer}.   
They demonstrate that several neural operator architectures can in fact be seen as CNFs. Depending on the architecture, the conditioning is local and/or global. By using the CNF interpretation, they also suggest an alternative implementation for the frequently used Fourier neural operator (FNO) that does not rely on regular grids, as in the standard setting, but on arbitrary points. 
Moreover, Catalani et al. \cite{CATALANI2026106929} note that their proposed MARIO framework is a generalization of the DeepONet architecture \cite{Lu_2021}. A similar observation has been made for the neural implicit flow framework \cite{JMLR:pan}, which combines a neural field with a hypernetwork processing external conditions. 
Mind that physics-informed neural networks (PINNs) \cite{RAISSI2019686} are neural fields as well, since they process point coordinates and map them to a certain physical quantity. They are especially useful when training data is not available. 

This paper is part of ongoing research for the development of data-driven surrogate models for the prediction of aircraft ditching loads and deformations. Ditching refers to emergency landing on water. Certification of a commercial aircraft requires ditching analysis, which can be based on simulations. A one-way coupling of the fluid-structure interaction is usually employed due to the high computational effort of the structural analysis \cite{schwarz:2024, Streckwall07}. To enable a two-way coupling while preserving computational speed, the long-term goal of this research is to include ML-based approximations of the deformations during the load analysis. First steps using ML to predict deformation-inducing ditching loads were investigated in previous works: Loads on the fuselage of a generic DLR-D150 aircraft were predicted using CAE-LSTM combinations \cite{schwarz:2024}. Subsequently,  the interpretability was improved by disentangling the latent space with a probabilistic $\beta$-variational autoencoder and deterministic alternatives \cite{schwarz:2025}. 
All examples of the datasets used in both previous studies are based on an identical discretization of the same fuselage.
This discretization is allowed to be more flexible in this work. For this purpose, a CNF approach
for dimensionality reduction is investigated, in which the latent variables are obtained using an autodecoding strategy.
The model is initially compared to a CAE-based model on  previously used datasets. Subsequently, the model is trained on a dataset containing results obtained from different discretizations. 
The ultimate goal of using a discretization independent approach is to consider different aircraft geometries with different discretizations within a single model. As a first step, we employ a single aircraft geometry with different discretizations in this work. 

The remainder of the paper is structured as follows: The CNF architecture and the training procedure are presented in Sec.~\ref{sec:model}. In Sec.~\ref{sec:results}, results on the two datasets are discussed and finally, conclusions are drawn in Sec.~\ref{sec:conclusions}.

\section{Conditional Neural Field}
\label{sec:model}
Neural fields predict the output for one coordinate at a time \cite{Xie:2022}. This way, they learn a continuous function on the coordinates, which is parameterized by the network parameters. To consider multiple data samples, the network can be conditioned to a latent vector using an autodecoder architecture \cite{Park_2019_CVPR}.  Different from an autoencoder, an autodecoder does not contain an encoder. The latent vectors are therefore considered to be trainable parameters that are optimized during training. 
The learned latent vectors can then be used for downstream tasks similar as in autoencoder-based models, e.g., to predict the temporal load evolution.

While this encoder-less construction is independent of the discretization and the numbers of points used for each sample, it also requires the optimization of the latent vector of a test sample at inference time \cite{Park_2019_CVPR}. As a result, the time for encoding a test sample is usually several orders of magnitude longer compared to autoencoder architectures.
In contrast, an advantage of this construction is that the output does not have to be calculated for the entire field if only individual points are of interest -- although the related computational efforts are often marginal in practice.

\subsection{Conditioning}
Conditioning allows a neural field to handle multiple different snap shots that are not necessarily obtained for the same discrete geometry.
There are different implementation strategies for conditioning \cite{Xie:2022}. A simple form of conditioning is the concatenation of the latent vector and the input coordinates. In the DeepSDF framework \cite{Park_2019_CVPR}, this concatenation is the input to the first hidden layer, and it is also connected to an intermediate layer with a skip connection.
The conditioning can be performed in a more complex way with a hypernetwork \cite{ha2017hypernetworks} that predicts the parameters of the neural field based on the latent vector \cite{sitzmann:2019,sitzmann:2020,JMLR:pan,yin2023continuous}, or with the feature-wise linear modulation (FiLM) \cite{Perez_Strub_deVries_Dumoulin_Courville_2018}. For the latter, the latent vector is transformed and fed into multiple layers of the network. 
Consider a layer output $\mathbf{x}\in\mathbb{R}^d$ and a latent vector $\mathbf{z}\in\mathbb{R}^m$. For functions $\boldsymbol{\gamma}, \boldsymbol{\beta}:\mathbb{R}^m \mapsto \mathbb{R}^d$, FiLM corresponds to an affine linear mapping
\begin{align}
    \boldsymbol{\gamma}(\mathbf{z})\odot\mathbf{x} + \boldsymbol{\beta}(\mathbf{z}),
\end{align}
where $\boldsymbol{\gamma}$ and $\boldsymbol{\beta}$ scale and shift $\mathbf{x}$, respectively.

Similar to \cite{serrano:2023,du:2024,CATALANI2026106929}, we employ FiLM in this work. 
Note that these three works only use the shift modulation $\boldsymbol{\beta}$. When $\boldsymbol{\beta}$ is a linear layer, this also corresponds to concatenating the layer output $\mathbf{x}$ and $\mathbf{z}$ \cite{dumoulin2018feature-wise}. For the CORAL framework, the authors report that using only the shift modulation led to more stable training and the same performance as using both shift and scale \cite{serrano:2023}. 
In our experiments, both options returned similar results as well. Therefore, we also only apply the shift modulation in this work. 

\subsection{Fourier Feature Mapping}
Neural networks are known for having a spectral bias, i.e., difficulties to learn high frequency content \cite{pmlr-v97-rahaman19a}. This issue can be addressed by using a sinusoidal representation network (SIREN) \cite{sitzmann:2020}, where periodic activation functions are applied. This approach is followed by the CORAL \cite{serrano:2023} and CoNFiLD \cite{du:2024} methods as well as the neural implicit flow framework \cite{JMLR:pan}. 
Alternatively, a positional encoding of the coordinates via a Fourier feature mapping (FFM) can be employed, which allows the network to learn different frequency components independently. 
Using neural tangent kernel (NTK) \cite{jacot:2018} theory, Tancik et al.\cite{tancik:2020} have shown that using an FFM overcomes the spectral bias and makes learning high frequency content easier. 
The mapping is defined as 
\begin{align}
\label{eq:FFM}
    \boldsymbol{\zeta}(\mathbf{v}) = [\cos(2\pi\mathbf{Bv}), \sin(2\pi\mathbf{Bv})]^T,
\end{align}
where $\mathbf{v}=(x,y,z)^\text{T}$ is the vector of input coordinates and the entries of $\mathbf{B}\in\mathbb{R}^{b\times 3}$ are sampled from $\mathcal{N}(0,\sigma^2)$ in this work. The dimension $b$ and the standard deviation $\sigma$ are hyperparameters whose values are given in Sec. \ref{sec:train}.

Like the MARIO framework \cite{CATALANI2026106929}, our model employs the FFM approach in this work.
To illustrate the benefits of FFM, training and validation loss curves for the case where the FFM was not used have been added to the Fig. \ref{fig:loss} in Sec. \ref{sec:results}.

\subsection{Full Model}
\label{sec:fullmodel}
\begin{figure*}[ht!]
    \centering
\includegraphics[width=0.7\linewidth]{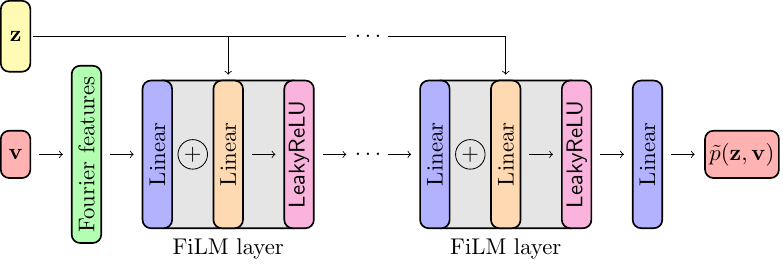}
    \caption{Scheme of the conditional neural field architecture.}
    \label{fig:cnf}
\end{figure*}

The full model is illustrated in Fig. \ref{fig:cnf}. 
For activation function $\alpha$, the input coordinate vector $\mathbf{v}$ and latent vector $\mathbf{z}$ are mapped to the pressure load prediction $\tilde{{p}}(\mathbf{z},\mathbf{v})$ via

\begin{align}
\begin{split}
    \mathbf{h}_0 &= 
    \boldsymbol{\zeta}(\mathbf{v})\\
    \mathbf{h}_i &= \alpha(\text{Linear}(\mathbf{h}_{i-1}) 
    + \boldsymbol{\beta}_i(\mathbf{z})), \quad 1 \le i \le N \\
    \tilde{{p}}(\mathbf{z},\mathbf{v}) &= \text{Linear}(\textbf{h}_N).
\end{split}
\end{align}
We implement the shift modulation $\boldsymbol{\beta}$ as a linear layer and use $\alpha=\textsf{LeakyReLU}$ with a slope parameter of $0.01$ in this work. 
Similar to the DeepSDF model \cite{Park_2019_CVPR}, we apply weight normalization \cite{salimans:2016}. Specifically, we apply it to the layers of the main network, i.e., the $\text{Linear}(\mathbf{h}_{i-1})$ layers. 
We use $N=6$ FiLM layers. 
Our CNF computes latent representations for each time step of a simulation. Thus, only the spatial and not the temporal coordinates are used as input to the CNF. This is consistent with the previously employed CAE-based models \cite{schwarz:2024, schwarz:2025}.

\subsection{Training}
\label{sec:train}
The model is trained with the mean squared error (MSE):
\begin{align}
    \text{MSE}(\mathbf{p}, \tilde{\mathbf{p}}) = \frac{1}{n}\Vert\mathbf{p-\tilde{\mathbf{p}}}\rVert^2_2,
    \quad \mathbf{p}\in \mathbb{R}^n,\tilde{\mathbf{p}}\in \mathbb{R}^n, 
\end{align}
where $\mathbf{p}$ and $\tilde{\mathbf{p}}$ are the true and reconstructed pressure values of a minibatch containing a total of $n$ spatial points.

The output values as well as the input coordinates are each normalized to the interval $[0,1]$ using $\min$-$\max$ normalization. The latent variables are initialized from $\mathcal{N}(0,0.1^2)$. In line with the previous work \cite{schwarz:2024, schwarz:2025}, we employ a latent space dimension of $m=10$. 
Regarding the FFM hyperparameters, we set $b=128$.
Too small [large] values of the standard deviation $\sigma$ can lead to underfitting [overfitting] \cite{tancik:2020}. 
In this work, $\sigma=10$ provided satisfactory results.
The CNF is trained for 1500 epochs.
For each of the two parameter groups, network parameters and latent vectors, 
an individual optimizer is defined and both are updated per minibatch. To this end, we use the Adam optimizer \cite{Kingma:2015} with a learning rate of $2\cdot 10^{-3}$ for both parameter groups. The minibatch size is set to 32. At inference time, the latent vectors for test examples are optimized while keeping the network parameters fixed.

Different learning strategies are conceivable, e.g., the meta-learning strategies used in \cite{serrano:2023, CATALANI2026106929}, based on the work of Zintgraf et al.\cite{pmlr-v97-zintgraf19a}. In this approach, the latent vectors are updated in an inner loop during training,  and the network parameters are updated subsequently in an outer loop using second-order gradients.
At inference time, this has the advantage that the latent vectors only have to be optimized for the same number of steps that are used in the inner loop during training. However, the computational cost for training is higher due to the second-order gradients. We initially also performed experiments using the meta-learning approach, but could not reach the same reconstruction quality as for the simpler first-order approach. Thus, 
we follow the first-order approach here, similar to of Park et al. \cite{Park_2019_CVPR} and Yin et al. \cite{yin2023continuous}, in which the number of epochs for the optimization of latent vectors at inference time is in the order of hundreds.

Similar as in \cite{Park_2019_CVPR,yin2023continuous, CATALANI2026106929}, we randomly sample a subset of spatial points for each example in a minibatch, containing 2000 spatial points. 
The error on the validation set can not be computed with a forward pass like in ordinary neural networks as the latent vectors have to be optimized. Optimizing the latent vector for all examples in the validation set in each epoch would therefore significantly prolong the training time. For that reason, the model and the latent vectors of the training set are stored at a few selected checkpoints throughout the entire training. The validation error is computed on all checkpoints after the training. 
Different from the training set, we use all available spatial points for the optimization of the latent vectors of the validation set as this lead to a more stable optimization in our experiments.
They are individually optimized for 200 epochs using Adam with a learning rate of $10^{-1}$. In our experiments, smaller learning rates often resulted in larger reconstruction errors.
The model is trained on two NVIDIA GeForce RTX 3090 GPUs in PyTorch \cite{NEURIPS2019_9015}. For validation and testing, a single GPU is employed.

\section{Results}
\label{sec:results}
\begin{figure}
    \centering
    \includegraphics[width=1\linewidth]{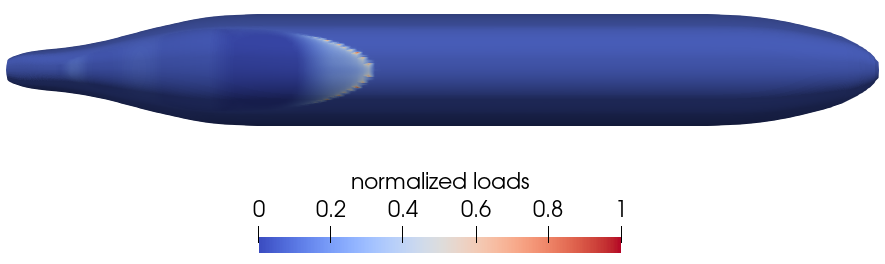}
    \caption{Exemplary normalized pressure loads on the D150 fuselage simulated with \emph{ditch}.}
    \label{fig:fuselage}
\end{figure}

The CNF model was trained and tested using two datasets containing simulated emergency water landing loads on the fuselage of a D150 aircraft. The data were obtained from simulations using different horizontal and vertical approach velocities to the free water surface. For this purpose, the in-house 3-degrees-of-freedom method \emph{ditch} \cite{bensch:2003} was used, which considers the dynamics of horizontal, vertical, and pitch motion. 
Further details on data generation and processing can be found in \cite{schwarz:2024}. Figure \ref{fig:fuselage} illustrates a view of the entire fuselage from below colored by contours of the normalized instantaneous pressure.

The aim of this study is twofold: firstly, to analyze the predictive capability of the CNF model in time-dependent dynamic processes (dataset A), and secondly, to analyze its predictive capability for heterogeneous spatial discretizations in the training and test data (dataset B). To reduce effort and complexity, the dynamic processes are analyzed using uniform spatial discretizations, and the influence of varying spatial discretizations is investigated without considering dynamic aspects.

For the first dataset, labeled dataset A, the uniform spatial discretization was already employed in two previous studies by the authors \cite{schwarz:2024, schwarz:2025}. In these works, CAE-LSTM combinations were compared to Koopman-based autoencoders for the spatio-temporal prediction of ditching loads \cite{schwarz:2024}, and the interpretability was improved by disentangling the latent space \cite{schwarz:2025}. 
The reconstruction accuracy as well as the ability to perform spatio-temporal predictions when paired with an LSTM network in the latent space are assessed. To this end, the results are compared to those obtained with a CAE.

For the second dataset, labeled dataset B, different discretizations are employed to demonstrate that the CNF can still reconstruct ditching loads accurately when trained on varying discretizations. As mentioned above, the second study is confined to static reconstructions, i.e., the temporal evolution is not considered.

\subsection{Dataset A}
\label{sec:dataset1}
The dataset contains a total of 373 ditching simulations with a D150 aircraft fuselage. 
These simulations differ in the initial horizontal and vertical velocities at the time of the first water impact, which range between approximately 67 and 87 m/s for the horizontal velocity, and 0.6 and 3.9 m/s for the vertical velocity, respectively. 

Each simulation consists of around 18--35 equidistant time steps of 0.1 s, which are represented by $128\times128$ sized images. These represent a portion of the fuselage, which in total is discretized using $150\times171$ spatial points.
The two spatial dimensions refer to the longitudinal ($x$) and circumferential coordinates. The circumferential coordinates contain contributions of both $y$- and $z$-coordinates and therefore three input coordinates $(x,y,z)$ to the CNF are considered. 

\begin{figure}[ht!]
    \centering
    \includegraphics[width=0.7\linewidth]{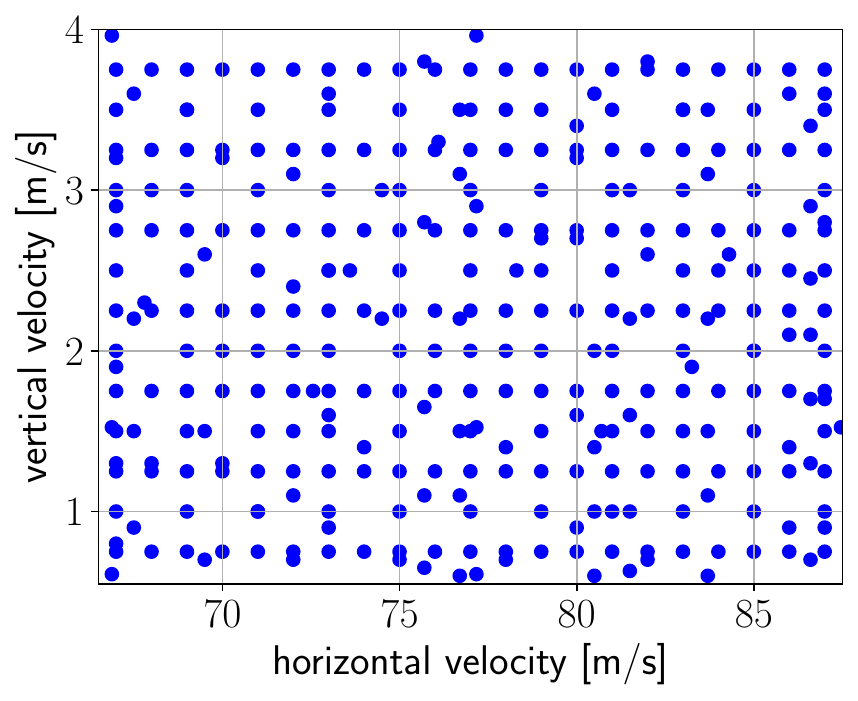}
    \caption{Velocity pairs for the training set of dataset A.
    }
    \label{fig:training_parameters_datasetA}
\end{figure}

The training set comprises 323 simulations, the distribution of the velocity pairs is shown in Fig.~\ref{fig:training_parameters_datasetA}.
For validation, data from 20 simulations is used and a test set containing results from 30 simulations serves to assess the spatio-temporal prediction accuracy (cf. Fig. \ref{fig:test_parameters_datasetB}).

\subsubsection{Dimension Reduction} 
\begin{figure*}[ht!]
    \centering
\includegraphics[width=1\linewidth]{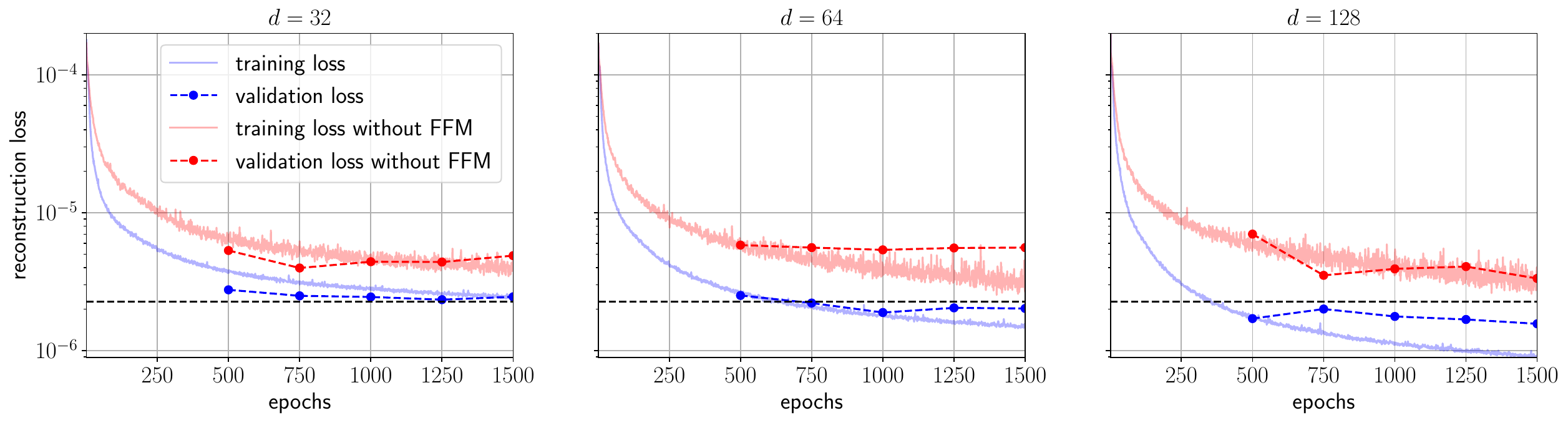}
    \caption{Reconstruction loss on the training and validation sets of dataset A for different dimensions $d$ of the CNF layers. The error on the validation set was computed on five checkpoints after the training. The dashed black line indicates the loss on the validation set for a CAE. Furthermore, loss curves are shown for the case where the Fourier Feature Mapping (FFM) has not been used to outline the benefit of FFM.}
    \label{fig:loss}
\end{figure*}

We compare results for different dimensions $d\in\{32,64,128\}$ of the CNF layers. 
Figure \ref{fig:loss} depicts the reconstruction loss on the training and validation set for the different dimensions $d$. 
The figure illustrates that a larger dimension $d$ leads to lower errors on the training set. Although this is the case for the validation set as well, the differences between the validation errors are much smaller for the smaller dimensions $d$. This leads to the training and validation loss curves for $d=32$ and $d=64$ being much closer to each other compared to $d=128$. 
For comparison, a CAE used in previous work \cite{schwarz:2025}  returns a validation error of $2.26\cdot10^{-6}$ when trained for 1500 epochs, cf. dashed horizontal line in Fig. \ref{fig:loss}. Note that in the previous work \cite{schwarz:2025} the CAE-based models were only trained for 500 epochs, as satisfactory results were achieved at that point. To ensure consistency with the CNF, we extend the CAE training time to 1500 epochs in this work.
As indicated in the left sub-figure, the validation error of the CAE is close to that of the CNF for $d=32$, and a slightly lower validation error is obtained for the two larger dimensions $d$.  
To illustrate the benefit of FFM, Fig. \ref{fig:loss} also includes results obtained without FFM that lead to a higher reconstruction loss 
under the same settings. As indicated by the dashed red lines, the validation error of the CNF without FFM is higher than that of the CAE for all tested dimensions $d$.

The training times, inference times for encoding and decoding, as well as the number of trainable parameters excluding latent vectors, are listed in Table \ref{tab:cnf_comparison}.
\begin{table*}[ht!]
    \centering
    \caption{Training times and inference times for encoding and decoding for different dimensions $d$ of the CNF in comparison to a previously used CAE \cite{schwarz:2025}. The CNF models are trained on two NVIDIA GeForce RTX 3090 GPUs. The CAE training time refers to 1500 epochs on a single NVIDIA GeForce RTX 3090 GPU and includes the calculation of the error on the validation set. Inference times are reported for a single GPU in all cases.}
    \begin{tabular}{lcccc}
    \hline
    \hline
    \multicolumn{1}{c}{Model} & \multicolumn{1}{c}{Training time [s]} & \multicolumn{1}{c}{Encoding inference time [s]} & \multicolumn{1}{c}{Decoding inference time [s]} & \multicolumn{1}{c}{Trainable parameters} \\
    \hline
     CNF(32) & 6595 & 1.91\phantom{000} & 0.00069 & \phantom{0}15841\\
     CNF(64) & 6525 & 1.92\phantom{000} & 0.00069 & \phantom{0}41921\\
     CNF(128) & 7030 & 2.05\phantom{000} & 0.00072 & 124801\\
     CAE & \phantom{0}870 & 0.00041 & 0.00039 & 134731\\
    \hline
    \hline
    \end{tabular}
    \label{tab:cnf_comparison}
\end{table*}
The different dimensions $d$ lead to fairly similar training times, although the largest value $d=128$ takes around 8\% more time than $d=32$ and $d=64$. 
In comparison to the CAE method, the training times for the CNF models are about eight times longer, with the CAE only being trained on one GPU and, like most neural networks, calculating the validation error already during the training process.
The times for encoding test samples are close to $2$ s for the three CNF models, with $d=128$ leading to around 7\% longer time than $d=32$ and $d=64$. {The validation set includes a total of 494 time steps, computing the validation error therefore requires around 945 s to 1015 s for each checkpoint, depending on $d$.}
The CAE, however, takes around four orders of magnitude less encoding inference time. For decoding, the inference times are in the same order of magnitude for all models, with the CAE again requiring the least time. 
{Note that when temporal predictions are performed in the latent space, typically only a few initial time steps need to be encoded but much more (predicted) time steps need to be decoded. The inference speed of the decoder is therefore usually more relevant for the use of a surrogate model.}
Considering the number of trainable parameters, all CNF models have less than the CAE. While for  $d=128$, the CNF has only around 7\% less parameters than the CAE, $d=64$ [$d=32$] lead to around 69\% [88\%] less parameters than the CAE. The memory efficiency of CNFs compared to grid-based methods has also been pointed out by Catalani et al. \cite{CATALANI2026106929}.
When the CNF has a similar amount of parameters as the previous CAE \cite{schwarz:2025} for $d=128$, a lower reconstruction error is obtained. Similarly, significantly less parameters are required to reach the same reconstruction quality. This comes, however, at the cost of longer training and inference times.

\subsubsection{Spatio-Temporal Predictions}
Unlike most previous engineering studies, this section aims to evaluate the performance of CNF in combination with an LSTM network in the latent space for predicting temporal evolutions.
The LSTM structure is based on the previous works \cite{schwarz:2024, schwarz:2025}, and comprises two LSTM layers with hidden state dimensions of 100, a linear output layer with $m=10$ neurons, and uses a time delay of three input time steps.
The network is again trained using the Adam optimizer with a learning rate of $10^{-3}$ and a minibatch size of 512 for 500 epochs.

To assess the accuracy of the spatio-temporal predictions, we use the same average error as in our previous studies \cite{schwarz:2024, schwarz:2025}. To this end, the models are trained five times to account for the randomness in the training process. From each trained CNF version, we use the latent vectors from the checkpoint with the lowest validation error to train the LSTM network, e.g., 
for the first version, the epochs 1250, 1000, and 1500 of the results shown in Fig. \ref{fig:loss} for CNF layer dimensions of 32, 64, and 128, respectively. We note that the validation errors obtained on the five checkpoints are not exactly the same over all five runs. 
To this end, predictions are made on all 30 test cases with each of the three trained CNF variants and the trained CAE. Normalized root mean squared errors (RMSEs) are computed for each test case, and the RMSEs are subsequently averaged over the five trained versions to obtain time-dependent errors for each test case. These are then time-averaged for each test case, and finally, an overall mean is calculated across all test cases.
The overall average errors are shown in Table \ref{tab:average_errors}.

\begin{table}[h!]
    \centering
    \caption{Average errors obtained by the different models combined with the same LSTM network for spatio-temporal predictions on 30 test cases.}
    \begin{tabular}{lr}
    \hline
    \hline
    Model & Average error \\
    \hline
     CNF(32) -- LSTM & 0.021\\
     CNF(64) -- LSTM & 0.022\\
     CNF(128) -- LSTM & 0.023\\
     CAE -- LSTM & 0.019\\
    \hline
    \hline
    \end{tabular}
    \label{tab:average_errors}
\end{table}

Although the reconstruction error of the CNF models is similar or lower than the CAE depending on the dimension $d$, the CNF-LSTM combination returns a slightly higher error level than the CAE-LSTM combination. 
A smaller CNF layer dimension $d$ generally leads to a lower average error in this setting, while obtaining higher reconstruction errors in Fig. \ref{fig:loss} when the temporal evolution is not considered. For $d=64$, the performance is similar to the uncorrelated autoencoder approach shown in our previous study\cite{schwarz:2025}, which corresponds to a CAE that aims to provide disentangled latent variables with a modified loss function.
It is important to note that the CNF for $d=128$ still performs better than the worst autoencoder models tested in previous studies\cite{schwarz:2024, schwarz:2025}. 

The reconstruction errors  on the validation set shown in Fig. \ref{fig:loss} for the stationary analyses no longer decrease significantly after epoch 500, especially for $d=32$ and $d=128$.
When the LSTM is trained with the CNF latent vectors corresponding to epoch 500, the average errors reported in Table \ref{tab:average_errors} remain nearly unchanged for $d=32$ and $d=128$, suggesting that the small training improvements observed after epoch 500 in Fig. \ref{fig:loss} do not have a significant influence on the time series prediction anymore. 
The training time could therefore be reduced to around $1/3$ of the time reported in Table \ref{tab:cnf_comparison} to achieve approximately the same accuracy for time series prediction. 
For $d=64$, however, the value increases slightly to 0.023. 
In the previous study, the CAE-LSTM combination also yielded a slightly higher average error value of 0.020 after 500 epochs of CAE training \cite{schwarz:2025}.

Setting the standard deviation $\sigma$ in the FFM in Eqn. \eqref{eq:FFM} too small [large] can lead to oversmoothed [noisy] results \cite{tancik:2020}. It can be expected that this also affects the obtained latent space. To investigate the related influence on the spatio-temporal predictions, we also performed experiments with a smaller value of $\sigma=1$ instead of $\sigma=10$. This lead to a small reduction of the average errors for $d=32$ and $d=128$, i.e., to 0.020 and 0.022 after 500 epochs of the CNF, respectively. For $d=32$, this approximately corresponds to the results obtained with a CAE after 500 epochs.
Values of 0.021 and 0.022 were obtained using the same procedure as for Table \ref{tab:average_errors}, i.e., using the epochs, where the lowest validation error is obtained over 1500 epochs. 
For $d=64$ however, this also resulted in small increments to 0.024 after 500 epochs and to 0.023 using the epochs where the lowest validation error is obtained over 1500 epochs. The experiments therefore did not provide a clear indication on the influence of $\sigma$ on the accuracy of the spatio-temporal predictions, at least not in the parameter range studied.

\subsection{Dataset B}
\begin{figure}[ht!]
    \centering
    \includegraphics[width=0.7\linewidth]{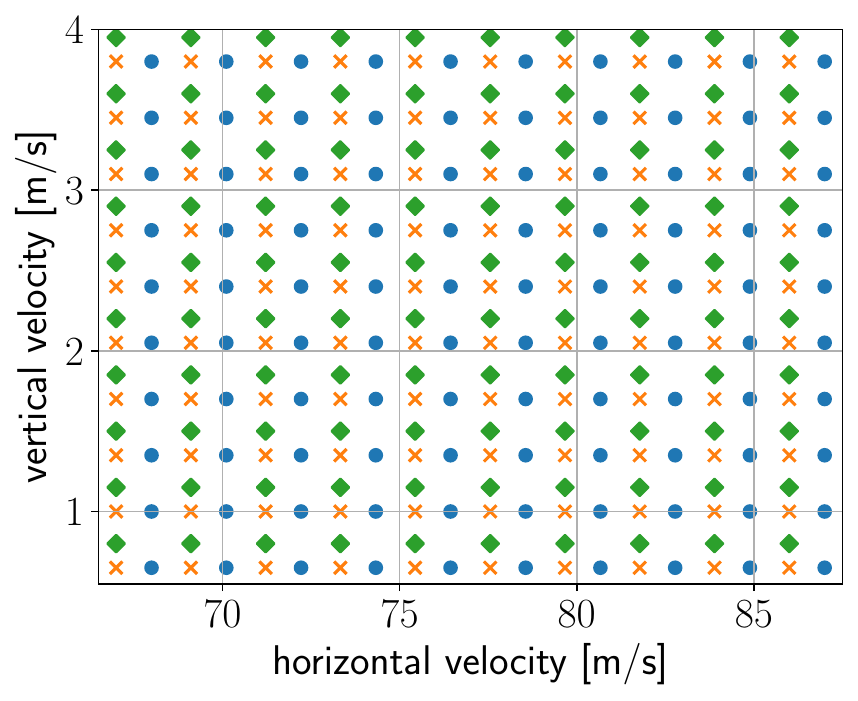}
    \caption{Velocity pairs for the training set of dataset B. Blue circles, orange crosses and green diamonds refer to sub-sets with 129, 150 and 170 points in longitudinal direction, respectively.
    }
    \label{fig:training_parameters_datasetB}
\end{figure}

The CNF is now trained using a dataset containing simulations with three different spatial discretizations of the same D150 hull. The training data of the second dataset is therefore not uniformly based on 150 $\times$ 171 discrete points, as is the case with dataset A, cf. Sec. \ref{sec:dataset1}. As in the first dataset, the number of points in the circumferential direction remains constant at 171. However, in the longitudinal direction, three different discretizations are used, based on 129, 150, and 170 discrete points, respectively. Each of these three longitudinal discretizations contributes to the training set with 100 simulations. The simulations again differ in the initial horizontal and vertical velocities.  
The training data space essentially corresponds to the first dataset and the distribution of the velocity pairs is depicted in Fig. \ref{fig:training_parameters_datasetB}, where the different colored shapes represent different sub-sets. Mind that the three sub-sets do not correspond to exactly the same distribution of velocity pairs as the dataset used in Sec. \ref{sec:dataset1}. However, the goal of this section is not to compare the accuracies of the trained models on both datasets, but rather to demonstrate that the CNF can naturally handle the second dataset and obtain satisfactory reconstruction quality.  
The validation set contains another 10 simulations for each discretization, and the input parameters are included in the operating range of the training set.
We consider the same portion of around 85\% of the total longitudinal coordinates as in Sec. \ref{sec:dataset1}, referring to 110, 128 and 145 considered points, respectively. To reduce training and inference time on the second dataset, we consider a smaller portion of the circumferential coordinates than in Sec. \ref{sec:dataset1}, i.e., the first 70 circumferential coordinates. 
Typically, circumferential coordinate positions beyond the 70th entry remain unwetted during the first impact and therefore very rarely contribute to the peak loads.
In line with that we also reduce the number of considered spatial points per example in a minibatch to 1000 during training.

\begin{figure}[h!]
    \centering
    \includegraphics[width=0.8\linewidth]{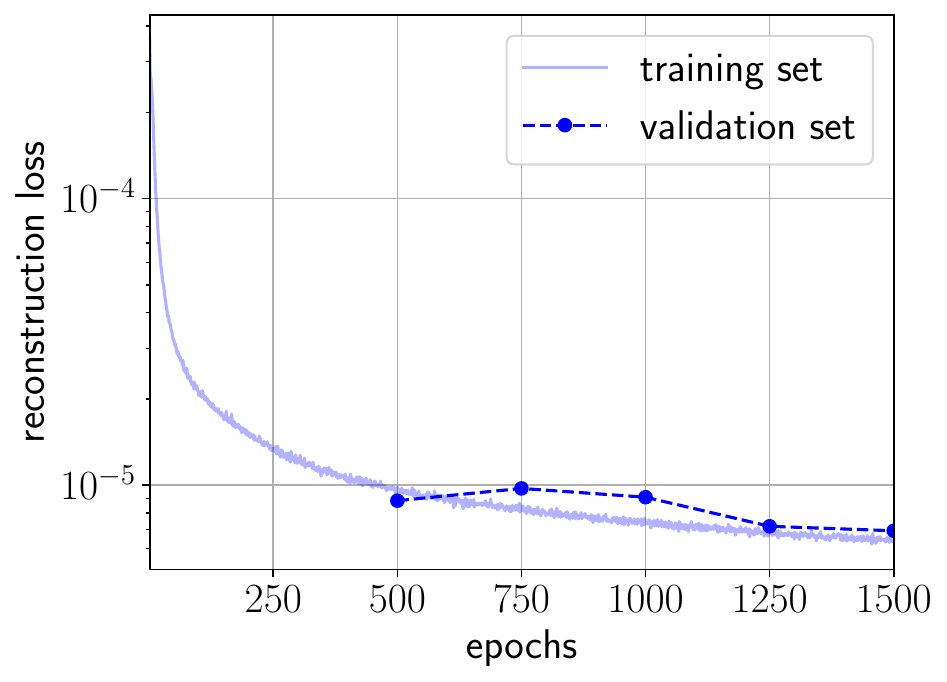}
    \caption{Reconstruction loss obtained for the CNF with dimension $d=32$ on the training and validation sets of dataset B.
    The error on the validation set was computed on five checkpoints after the training.}
    \label{fig:loss_irregular_data}
\end{figure}

Since the CAE algorithm cannot be applied to this dataset without further data processing, we restrict ourselves to the CNF and use the lowest tested hidden dimension $d=32$ from Sec. \ref{sec:dataset1} and $\sigma=10$ for the FFM.
Figure \ref{fig:loss_irregular_data} displays the loss on the training and validation sets. 

As seen in Sec. \ref{sec:dataset1}, the training and validation errors are again close together, suggesting that CNF can also generalize to unseen data for this dataset.

\subsubsection{Generalization to Unseen Discretizations}
\begin{figure}[ht!]
    \centering
    \includegraphics[width=0.7\linewidth]{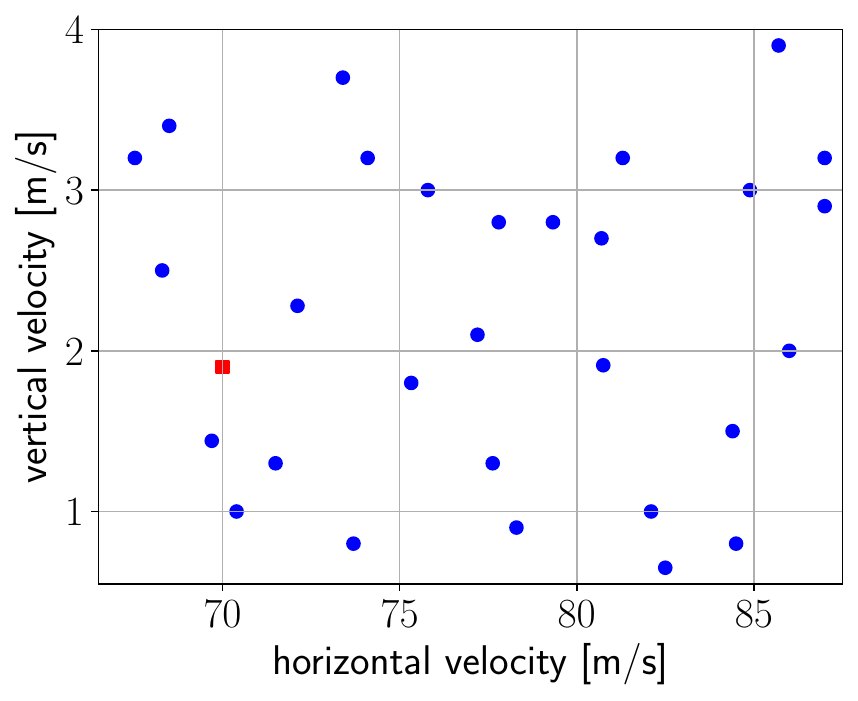}
    \caption{Velocity pairs for the test sets. Red square refers to the configuration, for which reconstructions of dataset B are shown in Fig.  \ref{fig:reconstructions_case29}. 
    }
    \label{fig:test_parameters_datasetB}
\end{figure}

\begin{figure*}
    \centering
    \includegraphics[width=0.92\linewidth]{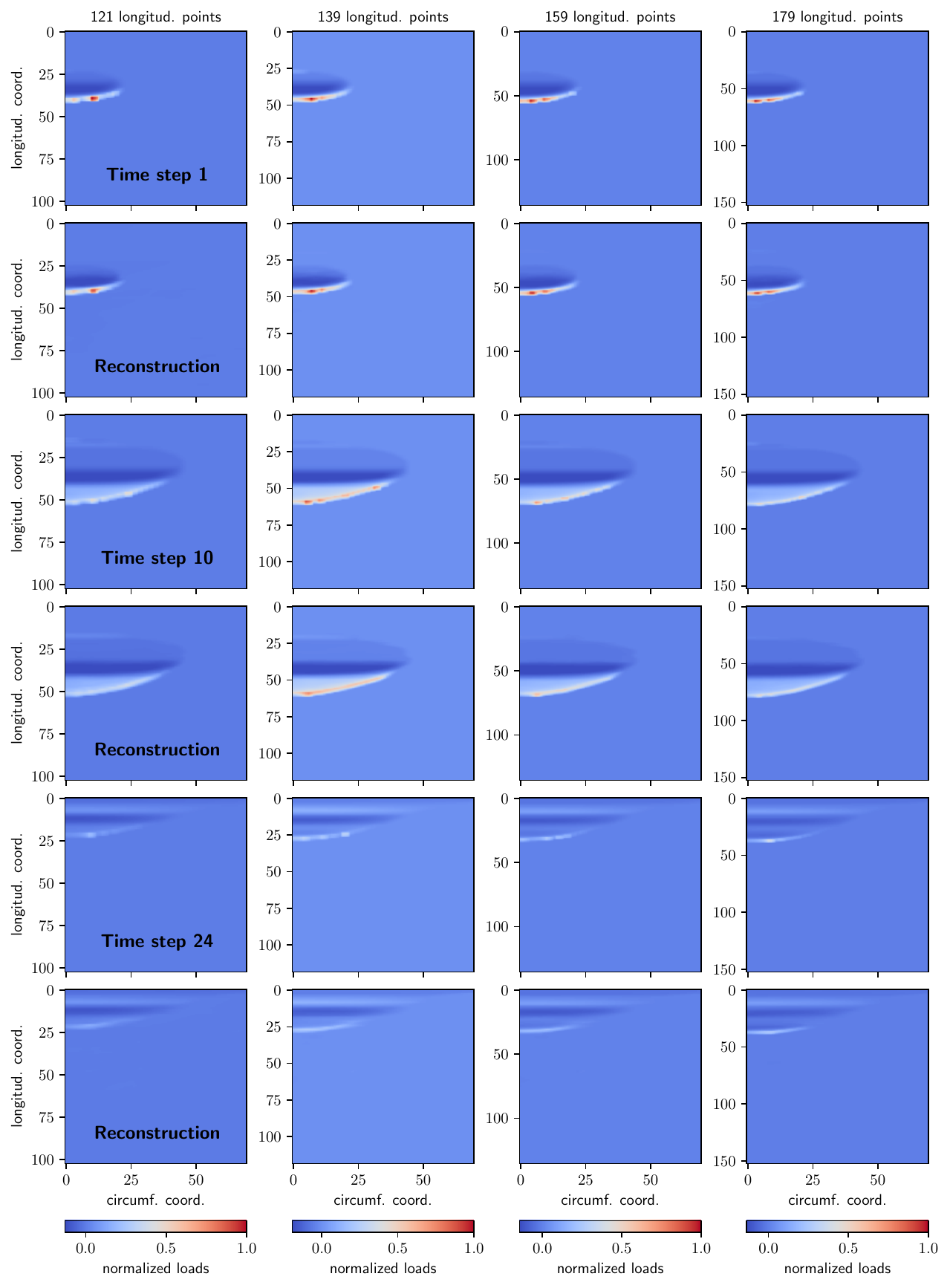}
    \caption{Simulated loads with the different unseen discretizations at three time steps of the test case and the corresponding CNF reconstructions.}
    \label{fig:reconstructions_case29}
\end{figure*}

We now test the model on discretizations that were not used during training. 
For this purpose, a test set composed of loads from four different discretizations is used. Specifically, 121, 139, 159 and 179 points in longitudinal direction are tested. Note that 121 and 179 points lie outside the range used for training while 139 and 159 lie inside. To evaluate the performance of the CNF, we again consider around 85\% of the longitudinal coordinates, i.e., 103, 119, 136 and 153 points, respectively.
For each discretization, 30 simulations are performed using the same input parameters for each discretization. These are again included in the operating range of the training set and essentially correspond to the test set in Sec. \ref{sec:dataset1}. The parameters are depicted in Fig. \ref{fig:test_parameters_datasetB}. The red colored parameter combination is employed for a qualitative assessment of the reconstruction quality returned by the different discretizations at the end of the section, cf. Fig. \ref{fig:reconstructions_case29}.
For comparison, we also perform these simulations for the three discretizations that were used during training. 
As for the previous simulations, a maximum time span of 7 s is considered for each test case \cite{schwarz:2024}. 
The period under consideration begins with the first impact, followed by a load intensive initial phase, and ends when the stresses fall below a lower threshold.
The total duration and the amount of considered time steps
depends on the approach conditions.
Mind that the different discretizations can lead to slightly different load histories, which can also influence the total duration and amount of time steps. Specifically, the 30 test cases result in a total of 807, 807, 810, 803, 792, 792 and 787 time steps for the longitudinal discretizations using 121, 129, 139, 150, 159, 170 and 179 points, respectively. This corresponds to a maximum difference of below 3\%. 
Results displayed in Fig. \ref{fig:loss_irregular_data}, motivate to employ the model from epoch 1500.

\begin{figure}[ht!]
    \centering
    \includegraphics[width=0.8\linewidth]{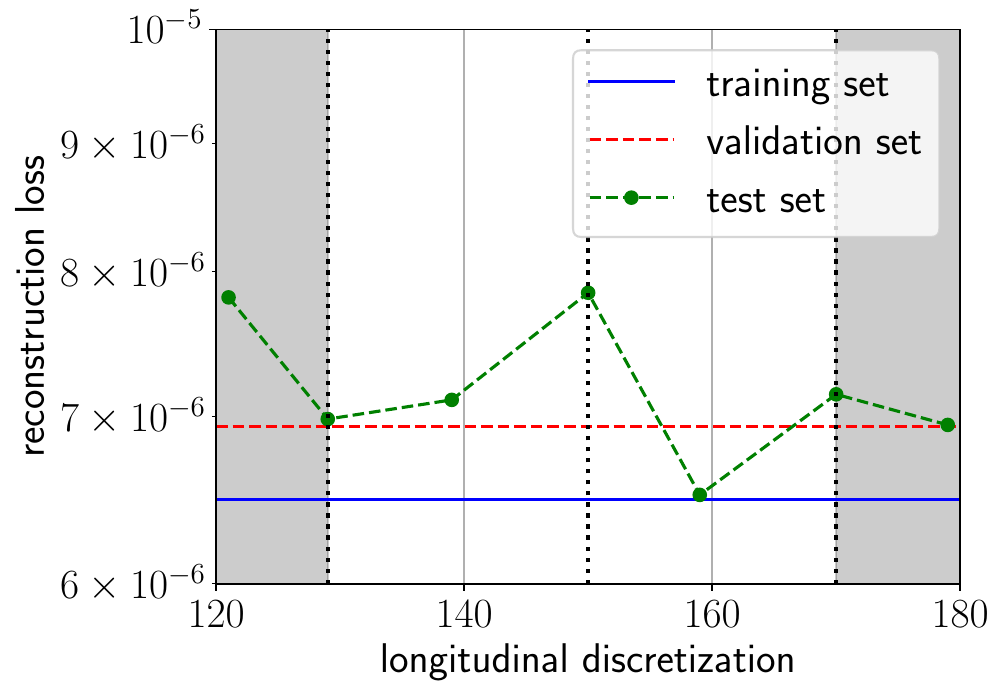}
    \caption{Reconstruction loss obtained for the CNF with dimension $d=32$ on the test set of dataset B. The errors on the training and validation sets refer to epoch 1500 in Fig. \ref{fig:loss_irregular_data}. The error on the test set is shown for each tested discretization. Dotted black lines indicate discretizations used during training. Shaded area indicates discretizations outside of range used during training.}
    \label{fig:test_loss_irregular_data}
\end{figure}
Figure \ref{fig:test_loss_irregular_data} depicts the reconstruction errors on the test sets for each discretization. The errors on the training and validation sets are also displayed for comparison.
In general, all errors are close together and within the range of what the studies shown above revealed.
The largest error is obtained for the longitudinal discretization using 150 points, followed by that with 121 points. The latter is outside of the range used for training while the former was used for training. Slightly lower errors are obtained for the discretizations with 170 and 139 points, followed by the ones using 129 and 180 points, respectively, which are similar to the validation error. The lowest error is returned for the discretization using 159 points and is close to the error on the training set. The discretization was not used during training but is inside the range of those used during training.
However, there is no obvious correlation between the error level and the type of discretization, i.e.  used or not used during training, or between error levels and the number of points in longitudinal direction. 
Moving further outside of the training range, the errors grow up to $1.06 \cdot 10^{-5}$ for a discretization with 98 points, and to $7.85\cdot10^{-6}$ [$8.23\cdot10^{-6}$] for a discretization with 197 [221] points.

Figure \ref{fig:reconstructions_case29} shows the pressure loads for exemplary time steps using the four tested discretizations that were not used during training, as well as their reconstructions for the respective discretizations.
The depicted time steps refer to the 1st, 10th and 24th time steps and cover three typical phases of a ditching process for an exemplary simulation with an initial horizontal and vertical velocity of 70 m/s and 1.9 m/s respectively, cf. Fig. \ref{fig:test_parameters_datasetB}. After the initial time step, in which the fuselage first touches the water, the wetted impact area widens at the 10th time step. During the final depicted time step, loads are primarily acting at the rear end of the fuselage. 
Mind that the same trained CNF is used for all four cases.
The results shown in the figure are of similar quality at the other test points, and strongly support the hypothesis that the CNF can accurately reconstruct the loads across different discretizations. Even with the 121-point discretization, which has the second largest error in Fig. \ref{fig:test_loss_irregular_data}, the quality of the reconstructions are satisfactory.

\section{Conclusion}
\label{sec:conclusions}
This paper presents a conditional neural field (CNF)-based surrogate model for ditching load prediction. 
Compared to a previously employed dimension reduction with a convolutional autoencoder (CAE) approach, the CNF approach is coordinate-based. It is therefore independent of the discretization and can be trained on datasets with varying discretizations.

On the first assessed dataset, all examples use the same discretization and the CNF achieved a similar reconstruction error as a CAE. Slightly lower errors were obtained for the larger CNF layer dimensions 64 and 128.
When paired with an LSTM network in the latent space, the CNF model, however, achieved slightly lower accuracy than the CAE on spatio-temporal ditching load predictions. The smallest tested layer dimension of 32 lead to the lowest error for the CNF in this case, while having around 88\% less parameters than the CAE.
The second dataset used consists of ditching loads that are trained for different discretizations and evaluated on yet other discretizations.
The results demonstrated that the CNF can generalize to discretizations that were not seen during training, which is a very useful feature for the engineering application of a surrogate model. 

General downsides of the CNF approach are the longer training and inference times compared to those of a CAE, especially for encoding test data. 
When data is available in the same discretization, CAE-based models thus have several advantages over the CNF approach. 
For more complicated datasets with heterogeneous discretizations, the CNF approach can be used without further data processing and therefore is the preferable choice. 

\section*{Acknowledgements}
H.S., P.P.L., \ and T.R.\ acknowledge support by the German Federal Ministry for Economic Affairs and Energy under aegis of the “Luftfahrtforschungsprogramm LuFo VI” project HYMNE (grant 20E2218A). 
This paper is a contribution to the research training group RTG 2583 on “Modeling, Simulation and Optimization of Fluid Dynamic Applications”  funded by the Deutsche Forschungsgemeinschaft (DFG). The authors acknowledge the support of TU Hamburg's Machine Learning in Engineering (MLE) initiative. 

\section*{Author Declarations}
\subsection*{Conflict of Interest}
The authors have no conflicts to disclose.

\subsection*{Author Contributions}
\textbf{Henning Schwarz}: Conceptualization (equal), Writing – original draft (equal), Writing – review \& editing (equal), Data curation (equal), Formal analysis (lead), Methodology (equal), Software (lead), Investigation (lead), Validation (lead), Visualization (lead).
\textbf{Pyei Phyo Lin}: Writing - Review \& Editing (equal), Data curation (equal), Methodology (equal).
\textbf{Jens-Peter M. Zemke}: Writing - Review \& Editing (equal), Conceptualization (equal), Methodology (equal).
\textbf{Thomas Rung}: Funding acquisition (lead), Writing – original draft (equal), Writing – review \& editing (equal), Conceptualization (equal), Methodology (equal), Project administration (lead), Resources (lead), Supervision (lead).

\section*{Data Availability}
The data that supports the findings of this study are available from the corresponding author upon reasonable request.

%
%

%


\bibliography{mybibliography.bib}

\begin{thebibliography}{45}%
\makeatletter
\providecommand \@ifxundefined [1]{%
 \@ifx{#1\undefined}
}%
\providecommand \@ifnum [1]{%
 \ifnum #1\expandafter \@firstoftwo
 \else \expandafter \@secondoftwo
 \fi
}%
\providecommand \@ifx [1]{%
 \ifx #1\expandafter \@firstoftwo
 \else \expandafter \@secondoftwo
 \fi
}%
\providecommand \natexlab [1]{#1}%
\providecommand \enquote  [1]{``#1''}%
\providecommand \bibnamefont  [1]{#1}%
\providecommand \bibfnamefont [1]{#1}%
\providecommand \citenamefont [1]{#1}%
\providecommand \href@noop [0]{\@secondoftwo}%
\providecommand \href [0]{\begingroup \@sanitize@url \@href}%
\providecommand \@href[1]{\@@startlink{#1}\@@href}%
\providecommand \@@href[1]{\endgroup#1\@@endlink}%
\providecommand \@sanitize@url [0]{\catcode `\\12\catcode `\$12\catcode
  `\&12\catcode `\#12\catcode `\^12\catcode `\_12\catcode `\%12\relax}%
\providecommand \@@startlink[1]{}%
\providecommand \@@endlink[0]{}%
\providecommand \url  [0]{\begingroup\@sanitize@url \@url }%
\providecommand \@url [1]{\endgroup\@href {#1}{\urlprefix }}%
\providecommand \urlprefix  [0]{URL }%
\providecommand \Eprint [0]{\href }%
\providecommand \doibase [0]{https://doi.org/}%
\providecommand \selectlanguage [0]{\@gobble}%
\providecommand \bibinfo  [0]{\@secondoftwo}%
\providecommand \bibfield  [0]{\@secondoftwo}%
\providecommand \translation [1]{[#1]}%
\providecommand \BibitemOpen [0]{}%
\providecommand \bibitemStop [0]{}%
\providecommand \bibitemNoStop [0]{.\EOS\space}%
\providecommand \EOS [0]{\spacefactor3000\relax}%
\providecommand \BibitemShut  [1]{\csname bibitem#1\endcsname}%
\let\auto@bib@innerbib\@empty
\bibitem [{\citenamefont {Lumley}(1967)}]{lumley1967structure}%
  \BibitemOpen
  \bibfield  {author} {\bibinfo {author} {\bibfnamefont {J.~L.}\ \bibnamefont
  {Lumley}},\ }\bibfield  {title} {\enquote {\bibinfo {title} {The structure of
  inhomogeneous turbulent flows},}\ }\href@noop {} {\bibfield  {journal}
  {\bibinfo  {journal} {Atmospheric turbulence and radio wave propagation}\ ,\
  \bibinfo {pages} {166--178}} (\bibinfo {year} {1967})}\BibitemShut {NoStop}%
\bibitem [{\citenamefont {Agostini}(2020)}]{Agostini20}%
  \BibitemOpen
  \bibfield  {author} {\bibinfo {author} {\bibfnamefont {L.}~\bibnamefont
  {Agostini}},\ }\bibfield  {title} {\enquote {\bibinfo {title} {Exploration
  and prediction of fluid dynamical systems using auto-encoder technology},}\
  }\href {https://doi.org/10.1063/5.0012906} {\bibfield  {journal} {\bibinfo
  {journal} {Physics of Fluids}\ }\textbf {\bibinfo {volume} {32}},\ \bibinfo
  {pages} {067103} (\bibinfo {year} {2020})}\BibitemShut {NoStop}%
\bibitem [{\citenamefont {Eivazi}\ \emph {et~al.}(2020)\citenamefont {Eivazi},
  \citenamefont {Veisi}, \citenamefont {Naderi},\ and\ \citenamefont
  {Esfahanian}}]{eivazi:2020}%
  \BibitemOpen
  \bibfield  {author} {\bibinfo {author} {\bibfnamefont {H.}~\bibnamefont
  {Eivazi}}, \bibinfo {author} {\bibfnamefont {H.}~\bibnamefont {Veisi}},
  \bibinfo {author} {\bibfnamefont {M.~H.}\ \bibnamefont {Naderi}},\ and\
  \bibinfo {author} {\bibfnamefont {V.}~\bibnamefont {Esfahanian}},\ }\bibfield
   {title} {\enquote {\bibinfo {title} {Deep neural networks for nonlinear
  model order reduction of unsteady flows},}\ }\href
  {https://doi.org/10.1063/5.0020526} {\bibfield  {journal} {\bibinfo
  {journal} {Physics of Fluids}\ }\textbf {\bibinfo {volume} {32}},\ \bibinfo
  {pages} {105104} (\bibinfo {year} {2020})}\BibitemShut {NoStop}%
\bibitem [{\citenamefont {Pache}\ and\ \citenamefont {Rung}(2022)}]{Pache22}%
  \BibitemOpen
  \bibfield  {author} {\bibinfo {author} {\bibfnamefont {R.}~\bibnamefont
  {Pache}}\ and\ \bibinfo {author} {\bibfnamefont {T.}~\bibnamefont {Rung}},\
  }\bibfield  {title} {\enquote {\bibinfo {title} {Data-driven surrogate
  modeling of aerodynamic forces on the superstructure of container vessels},}\
  }\href {https://doi.org/10.1080/19942060.2022.2044383} {\bibfield  {journal}
  {\bibinfo  {journal} {Engineering Applications of Computational Fluid
  Mechanics}\ }\textbf {\bibinfo {volume} {16}},\ \bibinfo {pages} {746--763}
  (\bibinfo {year} {2022})}\BibitemShut {NoStop}%
\bibitem [{\citenamefont {Eivazi}\ \emph {et~al.}(2022)\citenamefont {Eivazi},
  \citenamefont {Clainche}, \citenamefont {Hoyas},\ and\ \citenamefont
  {Vinuesa}}]{eivazi:2022}%
  \BibitemOpen
  \bibfield  {author} {\bibinfo {author} {\bibfnamefont {H.}~\bibnamefont
  {Eivazi}}, \bibinfo {author} {\bibfnamefont {S.~L.}\ \bibnamefont
  {Clainche}}, \bibinfo {author} {\bibfnamefont {S.}~\bibnamefont {Hoyas}},\
  and\ \bibinfo {author} {\bibfnamefont {R.}~\bibnamefont {Vinuesa}},\
  }\bibfield  {title} {\enquote {\bibinfo {title} {Towards extraction of
  orthogonal and parsimonious non-linear modes from turbulent flows},}\ }\href
  {https://doi.org/https://doi.org/10.1016/j.eswa.2022.117038} {\bibfield
  {journal} {\bibinfo  {journal} {Expert Systems with Applications}\ }\textbf
  {\bibinfo {volume} {202}},\ \bibinfo {pages} {117038} (\bibinfo {year}
  {2022})}\BibitemShut {NoStop}%
\bibitem [{\citenamefont {Kang}, \citenamefont {Yang},\ and\ \citenamefont
  {Yee}(2022)}]{kang:2022}%
  \BibitemOpen
  \bibfield  {author} {\bibinfo {author} {\bibfnamefont {Y.-E.}\ \bibnamefont
  {Kang}}, \bibinfo {author} {\bibfnamefont {S.}~\bibnamefont {Yang}},\ and\
  \bibinfo {author} {\bibfnamefont {K.}~\bibnamefont {Yee}},\ }\bibfield
  {title} {\enquote {\bibinfo {title} {{Physics-aware reduced-order modeling of
  transonic flow via $\beta$-variational autoencoder}},}\ }\href
  {https://doi.org/10.1063/5.0097740} {\bibfield  {journal} {\bibinfo
  {journal} {Physics of Fluids}\ }\textbf {\bibinfo {volume} {34}},\ \bibinfo
  {pages} {076103} (\bibinfo {year} {2022})}\BibitemShut {NoStop}%
\bibitem [{\citenamefont {{A. Solera-Rico, C. Sanmiguel Vila, M. Gómez-López,
  Y. Wang, A. Almashjary, S. Dawson, and R.
  Vinuesa}}(2024)}]{solera-rico:2024}%
  \BibitemOpen
  \bibfield  {author} {\bibinfo {author} {\bibnamefont {{A. Solera-Rico, C.
  Sanmiguel Vila, M. Gómez-López, Y. Wang, A. Almashjary, S. Dawson, and R.
  Vinuesa}}},\ }\bibfield  {title} {\enquote {\bibinfo {title}
  {$\beta$-variational autoencoders and transformers for reduced-order
  modelling of fluid flows},}\ }\href
  {https://doi.org/10.1038/s41467-024-45578-4} {\bibfield  {journal} {\bibinfo
  {journal} {Nature Communications}\ }\textbf {\bibinfo {volume} {15}}
  (\bibinfo {year} {2024}),\ 10.1038/s41467-024-45578-4}\BibitemShut {NoStop}%
\bibitem [{\citenamefont {Schwarz}\ \emph {et~al.}(2025)\citenamefont
  {Schwarz}, \citenamefont {\"{U}berr\"{u}ck}, \citenamefont {Zemke},\ and\
  \citenamefont {Rung}}]{schwarz:2024}%
  \BibitemOpen
  \bibfield  {author} {\bibinfo {author} {\bibfnamefont {H.}~\bibnamefont
  {Schwarz}}, \bibinfo {author} {\bibfnamefont {M.}~\bibnamefont
  {\"{U}berr\"{u}ck}}, \bibinfo {author} {\bibfnamefont {J.-P.~M.}\
  \bibnamefont {Zemke}},\ and\ \bibinfo {author} {\bibfnamefont
  {T.}~\bibnamefont {Rung}},\ }\bibfield  {title} {\enquote {\bibinfo {title}
  {Machine learning based prediction of ditching loads},}\ }\href
  {https://doi.org/10.2514/1.J064086} {\bibfield  {journal} {\bibinfo
  {journal} {AIAA Journal}\ }\textbf {\bibinfo {volume} {63}},\ \bibinfo
  {pages} {1835--1854} (\bibinfo {year} {2025})}\BibitemShut {NoStop}%
\bibitem [{\citenamefont {Loft}, \citenamefont {Schwarz},\ and\ \citenamefont
  {Rung}(2025)}]{loft:2025}%
  \BibitemOpen
  \bibfield  {author} {\bibinfo {author} {\bibfnamefont {M.}~\bibnamefont
  {Loft}}, \bibinfo {author} {\bibfnamefont {H.}~\bibnamefont {Schwarz}},\ and\
  \bibinfo {author} {\bibfnamefont {T.}~\bibnamefont {Rung}},\ }\bibfield
  {title} {\enquote {\bibinfo {title} {Data-driven pressure field prediction
  for ships in regular sea states},}\ }\href
  {https://doi.org/10.1080/19942060.2025.2553337} {\bibfield  {journal}
  {\bibinfo  {journal} {Engineering Applications of Computational Fluid
  Mechanics}\ }\textbf {\bibinfo {volume} {19}},\ \bibinfo {pages} {2553337}
  (\bibinfo {year} {2025})}\BibitemShut {NoStop}%
\bibitem [{\citenamefont {Hochreiter}\ and\ \citenamefont
  {Schmidhuber}(1997)}]{hochreiter:1997}%
  \BibitemOpen
  \bibfield  {author} {\bibinfo {author} {\bibfnamefont {S.}~\bibnamefont
  {Hochreiter}}\ and\ \bibinfo {author} {\bibfnamefont {J.}~\bibnamefont
  {Schmidhuber}},\ }\bibfield  {title} {\enquote {\bibinfo {title} {Long
  short-term memory},}\ }\href {https://doi.org/10.1162/neco.1997.9.8.1735}
  {\bibfield  {journal} {\bibinfo  {journal} {Neural Computation}\ }\textbf
  {\bibinfo {volume} {9.8}},\ \bibinfo {pages} {1735–1780} (\bibinfo {year}
  {1997})}\BibitemShut {NoStop}%
\bibitem [{\citenamefont {{A. Vaswani, N. Shazeer, N. Parmar, J. Uszkoreit, L.
  Jones, A. N. Gomez, L. Kaiser, and I. Polosukhin}}(2017)}]{vaswani:2017}%
  \BibitemOpen
  \bibfield  {author} {\bibinfo {author} {\bibnamefont {{A. Vaswani, N.
  Shazeer, N. Parmar, J. Uszkoreit, L. Jones, A. N. Gomez, L. Kaiser, and I.
  Polosukhin}}},\ }\bibfield  {title} {\enquote {\bibinfo {title} {Attention is
  all you need},}\ }in\ \href
  {https://proceedings.neurips.cc/paper/2017/file/3f5ee243547dee91fbd053c1c4a845aa-Paper.pdf}
  {\emph {\bibinfo {booktitle} {Advances in Neural Information Processing
  Systems}}}\ (\bibinfo {year} {2017})\BibitemShut {NoStop}%
\bibitem [{\citenamefont {Wu}\ \emph {et~al.}(2021)\citenamefont {Wu},
  \citenamefont {Gong}, \citenamefont {Pan}, \citenamefont {Qiu}, \citenamefont
  {Feng},\ and\ \citenamefont {Pain}}]{wu:2021}%
  \BibitemOpen
  \bibfield  {author} {\bibinfo {author} {\bibfnamefont {P.}~\bibnamefont
  {Wu}}, \bibinfo {author} {\bibfnamefont {S.}~\bibnamefont {Gong}}, \bibinfo
  {author} {\bibfnamefont {K.}~\bibnamefont {Pan}}, \bibinfo {author}
  {\bibfnamefont {F.}~\bibnamefont {Qiu}}, \bibinfo {author} {\bibfnamefont
  {W.}~\bibnamefont {Feng}},\ and\ \bibinfo {author} {\bibfnamefont
  {C.}~\bibnamefont {Pain}},\ }\bibfield  {title} {\enquote {\bibinfo {title}
  {Reduced order model using convolutional auto-encoder with self-attention},}\
  }\href {https://doi.org/10.1063/5.0051155} {\bibfield  {journal} {\bibinfo
  {journal} {Physics of Fluids}\ }\textbf {\bibinfo {volume} {33}},\ \bibinfo
  {pages} {077107} (\bibinfo {year} {2021})}\BibitemShut {NoStop}%
\bibitem [{\citenamefont {Hemmasian}\ and\ \citenamefont
  {Barati~Farimani}(2023)}]{Hemmasian:2023}%
  \BibitemOpen
  \bibfield  {author} {\bibinfo {author} {\bibfnamefont {A.}~\bibnamefont
  {Hemmasian}}\ and\ \bibinfo {author} {\bibfnamefont {A.}~\bibnamefont
  {Barati~Farimani}},\ }\bibfield  {title} {\enquote {\bibinfo {title}
  {{Reduced-order modeling of fluid flows with transformers}},}\ }\href
  {https://doi.org/10.1063/5.0151515} {\bibfield  {journal} {\bibinfo
  {journal} {Physics of Fluids}\ }\textbf {\bibinfo {volume} {35}},\ \bibinfo
  {pages} {057126} (\bibinfo {year} {2023})}\BibitemShut {NoStop}%
\bibitem [{\citenamefont {Maram}\ \emph {et~al.}(2026)\citenamefont {Maram},
  \citenamefont {Bletsos}, \citenamefont {Nguyen}, \citenamefont {Hassan},
  \citenamefont {Palm},\ and\ \citenamefont
  {Rung}}]{maram2026adjointbasedshapeoptimizationship}%
  \BibitemOpen
  \bibfield  {author} {\bibinfo {author} {\bibfnamefont {M.~A.}\ \bibnamefont
  {Maram}}, \bibinfo {author} {\bibfnamefont {G.}~\bibnamefont {Bletsos}},
  \bibinfo {author} {\bibfnamefont {T.~T.}\ \bibnamefont {Nguyen}}, \bibinfo
  {author} {\bibfnamefont {A.}~\bibnamefont {Hassan}}, \bibinfo {author}
  {\bibfnamefont {M.}~\bibnamefont {Palm}},\ and\ \bibinfo {author}
  {\bibfnamefont {T.}~\bibnamefont {Rung}},\ }\href
  {https://arxiv.org/abs/2602.14907} {\enquote {\bibinfo {title} {Adjoint-based
  shape optimization of a ship hull using a conditional variational autoencoder
  (cvae) assisted propulsion surrogate model},}\ } (\bibinfo {year} {2026}),\
  \Eprint {https://arxiv.org/abs/2602.14907} {arXiv:2602.14907
  [physics.flu-dyn]} \BibitemShut {NoStop}%
\bibitem [{\citenamefont {Pfaff}\ \emph {et~al.}(2021)\citenamefont {Pfaff},
  \citenamefont {Fortunato}, \citenamefont {Sanchez-Gonzalez},\ and\
  \citenamefont {Battaglia}}]{pfaff2021learning}%
  \BibitemOpen
  \bibfield  {author} {\bibinfo {author} {\bibfnamefont {T.}~\bibnamefont
  {Pfaff}}, \bibinfo {author} {\bibfnamefont {M.}~\bibnamefont {Fortunato}},
  \bibinfo {author} {\bibfnamefont {A.}~\bibnamefont {Sanchez-Gonzalez}},\ and\
  \bibinfo {author} {\bibfnamefont {P.}~\bibnamefont {Battaglia}},\ }\bibfield
  {title} {\enquote {\bibinfo {title} {Learning mesh-based simulation with
  graph networks},}\ }in\ \href@noop {} {\emph {\bibinfo {booktitle}
  {International Conference on Learning Representations}}}\ (\bibinfo {year}
  {2021})\BibitemShut {NoStop}%
\bibitem [{\citenamefont {Brandstetter}, \citenamefont {Worrall},\ and\
  \citenamefont {Welling}(2022)}]{brandstetter2022message}%
  \BibitemOpen
  \bibfield  {author} {\bibinfo {author} {\bibfnamefont {J.}~\bibnamefont
  {Brandstetter}}, \bibinfo {author} {\bibfnamefont {D.~E.}\ \bibnamefont
  {Worrall}},\ and\ \bibinfo {author} {\bibfnamefont {M.}~\bibnamefont
  {Welling}},\ }\bibfield  {title} {\enquote {\bibinfo {title} {Message passing
  neural {PDE} solvers},}\ }in\ \href@noop {} {\emph {\bibinfo {booktitle}
  {International Conference on Learning Representations}}}\ (\bibinfo {year}
  {2022})\BibitemShut {NoStop}%
\bibitem [{\citenamefont {Barwey}\ \emph {et~al.}(2023)\citenamefont {Barwey},
  \citenamefont {Shankar}, \citenamefont {Viswanathan},\ and\ \citenamefont
  {Maulik}}]{BARWEY2023112537}%
  \BibitemOpen
  \bibfield  {author} {\bibinfo {author} {\bibfnamefont {S.}~\bibnamefont
  {Barwey}}, \bibinfo {author} {\bibfnamefont {V.}~\bibnamefont {Shankar}},
  \bibinfo {author} {\bibfnamefont {V.}~\bibnamefont {Viswanathan}},\ and\
  \bibinfo {author} {\bibfnamefont {R.}~\bibnamefont {Maulik}},\ }\bibfield
  {title} {\enquote {\bibinfo {title} {Multiscale graph neural network
  autoencoders for interpretable scientific machine learning},}\ }\href
  {https://doi.org/https://doi.org/10.1016/j.jcp.2023.112537} {\bibfield
  {journal} {\bibinfo  {journal} {Journal of Computational Physics}\ }\textbf
  {\bibinfo {volume} {495}},\ \bibinfo {pages} {112537} (\bibinfo {year}
  {2023})}\BibitemShut {NoStop}%
\bibitem [{\citenamefont {Franco}\ \emph {et~al.}(2023)\citenamefont {Franco},
  \citenamefont {Fresca}, \citenamefont {Tombari},\ and\ \citenamefont
  {Manzoni}}]{franco:2023}%
  \BibitemOpen
  \bibfield  {author} {\bibinfo {author} {\bibfnamefont {N.~R.}\ \bibnamefont
  {Franco}}, \bibinfo {author} {\bibfnamefont {S.}~\bibnamefont {Fresca}},
  \bibinfo {author} {\bibfnamefont {F.}~\bibnamefont {Tombari}},\ and\ \bibinfo
  {author} {\bibfnamefont {A.}~\bibnamefont {Manzoni}},\ }\bibfield  {title}
  {\enquote {\bibinfo {title} {Deep learning-based surrogate models for
  parametrized pdes: Handling geometric variability through graph neural
  networks},}\ }\href@noop {} {\bibfield  {journal} {\bibinfo  {journal}
  {Chaos: An Interdisciplinary Journal of Nonlinear Science}\ }\textbf
  {\bibinfo {volume} {33}},\ \bibinfo {pages} {123121} (\bibinfo {year}
  {2023})}\BibitemShut {NoStop}%
\bibitem [{\citenamefont {Serrano}\ \emph {et~al.}(2023)\citenamefont
  {Serrano}, \citenamefont {Le~Boudec}, \citenamefont
  {Kassa\"{\i}~Koupa\"{\i}}, \citenamefont {Wang}, \citenamefont {Yin},
  \citenamefont {Vittaut},\ and\ \citenamefont {Gallinari}}]{serrano:2023}%
  \BibitemOpen
  \bibfield  {author} {\bibinfo {author} {\bibfnamefont {L.}~\bibnamefont
  {Serrano}}, \bibinfo {author} {\bibfnamefont {L.}~\bibnamefont {Le~Boudec}},
  \bibinfo {author} {\bibfnamefont {A.}~\bibnamefont
  {Kassa\"{\i}~Koupa\"{\i}}}, \bibinfo {author} {\bibfnamefont {T.~X.}\
  \bibnamefont {Wang}}, \bibinfo {author} {\bibfnamefont {Y.}~\bibnamefont
  {Yin}}, \bibinfo {author} {\bibfnamefont {J.-N.}\ \bibnamefont {Vittaut}},\
  and\ \bibinfo {author} {\bibfnamefont {P.}~\bibnamefont {Gallinari}},\
  }\bibfield  {title} {\enquote {\bibinfo {title} {Operator learning with
  neural fields: Tackling pdes on general geometries},}\ }in\ \href
  {https://proceedings.neurips.cc/paper_files/paper/2023/file/df54302388bbc145aacaa1a54a4a5933-Paper-Conference.pdf}
  {\emph {\bibinfo {booktitle} {Advances in Neural Information Processing
  Systems}}},\ Vol.~\bibinfo {volume} {36},\ \bibinfo {editor} {edited by\
  \bibinfo {editor} {\bibfnamefont {A.}~\bibnamefont {Oh}}, \bibinfo {editor}
  {\bibfnamefont {T.}~\bibnamefont {Naumann}}, \bibinfo {editor} {\bibfnamefont
  {A.}~\bibnamefont {Globerson}}, \bibinfo {editor} {\bibfnamefont
  {K.}~\bibnamefont {Saenko}}, \bibinfo {editor} {\bibfnamefont
  {M.}~\bibnamefont {Hardt}},\ and\ \bibinfo {editor} {\bibfnamefont
  {S.}~\bibnamefont {Levine}}}\ (\bibinfo  {publisher} {Curran Associates,
  Inc.},\ \bibinfo {year} {2023})\ pp.\ \bibinfo {pages}
  {70581--70611}\BibitemShut {NoStop}%
\bibitem [{\citenamefont {Xie}\ \emph {et~al.}()\citenamefont {Xie},
  \citenamefont {Takikawa}, \citenamefont {Saito}, \citenamefont {Litany},
  \citenamefont {Yan}, \citenamefont {Khan}, \citenamefont {Tombari},
  \citenamefont {Tompkin}, \citenamefont {Sitzmann},\ and\ \citenamefont
  {Sridhar}}]{Xie:2022}%
  \BibitemOpen
  \bibfield  {author} {\bibinfo {author} {\bibfnamefont {Y.}~\bibnamefont
  {Xie}}, \bibinfo {author} {\bibfnamefont {T.}~\bibnamefont {Takikawa}},
  \bibinfo {author} {\bibfnamefont {S.}~\bibnamefont {Saito}}, \bibinfo
  {author} {\bibfnamefont {O.}~\bibnamefont {Litany}}, \bibinfo {author}
  {\bibfnamefont {S.}~\bibnamefont {Yan}}, \bibinfo {author} {\bibfnamefont
  {N.}~\bibnamefont {Khan}}, \bibinfo {author} {\bibfnamefont {F.}~\bibnamefont
  {Tombari}}, \bibinfo {author} {\bibfnamefont {J.}~\bibnamefont {Tompkin}},
  \bibinfo {author} {\bibfnamefont {V.}~\bibnamefont {Sitzmann}},\ and\
  \bibinfo {author} {\bibfnamefont {S.}~\bibnamefont {Sridhar}},\ }\bibfield
  {title} {\enquote {\bibinfo {title} {Neural fields in visual computing and
  beyond},}\ }\href {https://doi.org/https://doi.org/10.1111/cgf.14505}
  {\bibfield  {journal} {\bibinfo  {journal} {Computer Graphics Forum}\
  }\textbf {\bibinfo {volume} {41}},\ \bibinfo {pages} {641--676}}\BibitemShut
  {NoStop}%
\bibitem [{\citenamefont {Park}\ \emph {et~al.}(2019)\citenamefont {Park},
  \citenamefont {Florence}, \citenamefont {Straub}, \citenamefont {Newcombe},\
  and\ \citenamefont {Lovegrove}}]{Park_2019_CVPR}%
  \BibitemOpen
  \bibfield  {author} {\bibinfo {author} {\bibfnamefont {J.~J.}\ \bibnamefont
  {Park}}, \bibinfo {author} {\bibfnamefont {P.}~\bibnamefont {Florence}},
  \bibinfo {author} {\bibfnamefont {J.}~\bibnamefont {Straub}}, \bibinfo
  {author} {\bibfnamefont {R.}~\bibnamefont {Newcombe}},\ and\ \bibinfo
  {author} {\bibfnamefont {S.}~\bibnamefont {Lovegrove}},\ }\bibfield  {title}
  {\enquote {\bibinfo {title} {Deepsdf: Learning continuous signed distance
  functions for shape representation},}\ }in\ \href@noop {} {\emph {\bibinfo
  {booktitle} {Proceedings of the IEEE/CVF Conference on Computer Vision and
  Pattern Recognition (CVPR)}}}\ (\bibinfo {year} {2019})\BibitemShut {NoStop}%
\bibitem [{\citenamefont {Yin}\ \emph {et~al.}(2023)\citenamefont {Yin},
  \citenamefont {Kirchmeyer}, \citenamefont {Franceschi}, \citenamefont
  {Rakotomamonjy},\ and\ \citenamefont {Gallinari}}]{yin2023continuous}%
  \BibitemOpen
  \bibfield  {author} {\bibinfo {author} {\bibfnamefont {Y.}~\bibnamefont
  {Yin}}, \bibinfo {author} {\bibfnamefont {M.}~\bibnamefont {Kirchmeyer}},
  \bibinfo {author} {\bibfnamefont {J.-Y.}\ \bibnamefont {Franceschi}},
  \bibinfo {author} {\bibfnamefont {A.}~\bibnamefont {Rakotomamonjy}},\ and\
  \bibinfo {author} {\bibfnamefont {P.}~\bibnamefont {Gallinari}},\ }\bibfield
  {title} {\enquote {\bibinfo {title} {Continuous {PDE} dynamics forecasting
  with implicit neural representations},}\ }in\ \href@noop {} {\emph {\bibinfo
  {booktitle} {The Eleventh International Conference on Learning
  Representations}}}\ (\bibinfo {year} {2023})\BibitemShut {NoStop}%
\bibitem [{\citenamefont {Du}\ \emph {et~al.}(2024)\citenamefont {Du},
  \citenamefont {Parikh}, \citenamefont {Fan}, \citenamefont {Liu},\ and\
  \citenamefont {Wang}}]{du:2024}%
  \BibitemOpen
  \bibfield  {author} {\bibinfo {author} {\bibfnamefont {P.}~\bibnamefont
  {Du}}, \bibinfo {author} {\bibfnamefont {M.~H.}\ \bibnamefont {Parikh}},
  \bibinfo {author} {\bibfnamefont {X.}~\bibnamefont {Fan}}, \bibinfo {author}
  {\bibfnamefont {X.-Y.}\ \bibnamefont {Liu}},\ and\ \bibinfo {author}
  {\bibfnamefont {J.-X.}\ \bibnamefont {Wang}},\ }\bibfield  {title} {\enquote
  {\bibinfo {title} {Conditional neural field latent diffusion model for
  generating spatiotemporal turbulence},}\ }\href
  {https://doi.org/10.1038/s41467-024-54712-1} {\bibfield  {journal} {\bibinfo
  {journal} {Nature Communications}\ }\textbf {\bibinfo {volume} {15}}
  (\bibinfo {year} {2024}),\ 10.1038/s41467-024-54712-1}\BibitemShut {NoStop}%
\bibitem [{\citenamefont {Guo}\ \emph {et~al.}(2026)\citenamefont {Guo},
  \citenamefont {Du}, \citenamefont {Fan}, \citenamefont {Li},\ and\
  \citenamefont {Wang}}]{guo:2026}%
  \BibitemOpen
  \bibfield  {author} {\bibinfo {author} {\bibfnamefont {J.}~\bibnamefont
  {Guo}}, \bibinfo {author} {\bibfnamefont {P.}~\bibnamefont {Du}}, \bibinfo
  {author} {\bibfnamefont {X.}~\bibnamefont {Fan}}, \bibinfo {author}
  {\bibfnamefont {Y.}~\bibnamefont {Li}},\ and\ \bibinfo {author}
  {\bibfnamefont {J.-X.}\ \bibnamefont {Wang}},\ }\bibfield  {title} {\enquote
  {\bibinfo {title} {Conditional neural field for spatial dimension reduction
  of turbulence data: A comparison study},}\ }\href
  {https://doi.org/10.1063/5.0310238} {\bibfield  {journal} {\bibinfo
  {journal} {Physics of Fluids}\ }\textbf {\bibinfo {volume} {38}},\ \bibinfo
  {pages} {025153} (\bibinfo {year} {2026})}\BibitemShut {NoStop}%
\bibitem [{\citenamefont {Catalani}\ \emph {et~al.}(2026)\citenamefont
  {Catalani}, \citenamefont {Fesquet}, \citenamefont {Bertrand}, \citenamefont
  {Tost}, \citenamefont {Bauerheim},\ and\ \citenamefont
  {Morlier}}]{CATALANI2026106929}%
  \BibitemOpen
  \bibfield  {author} {\bibinfo {author} {\bibfnamefont {G.}~\bibnamefont
  {Catalani}}, \bibinfo {author} {\bibfnamefont {J.}~\bibnamefont {Fesquet}},
  \bibinfo {author} {\bibfnamefont {X.}~\bibnamefont {Bertrand}}, \bibinfo
  {author} {\bibfnamefont {F.}~\bibnamefont {Tost}}, \bibinfo {author}
  {\bibfnamefont {M.}~\bibnamefont {Bauerheim}},\ and\ \bibinfo {author}
  {\bibfnamefont {J.}~\bibnamefont {Morlier}},\ }\bibfield  {title} {\enquote
  {\bibinfo {title} {Towards scalable surrogate models based on neural fields
  for large scale aerodynamic simulations},}\ }\href
  {https://doi.org/https://doi.org/10.1016/j.compfluid.2025.106929} {\bibfield
  {journal} {\bibinfo  {journal} {Computers \& Fluids}\ }\textbf {\bibinfo
  {volume} {306}},\ \bibinfo {pages} {106929} (\bibinfo {year}
  {2026})}\BibitemShut {NoStop}%
\bibitem [{\citenamefont {Li}\ \emph {et~al.}(2021)\citenamefont {Li},
  \citenamefont {Kovachki}, \citenamefont {Azizzadenesheli}, \citenamefont
  {Liu}, \citenamefont {Bhattacharya}, \citenamefont {Stuart},\ and\
  \citenamefont {Anandkumar}}]{li2021fourier}%
  \BibitemOpen
  \bibfield  {author} {\bibinfo {author} {\bibfnamefont {Z.}~\bibnamefont
  {Li}}, \bibinfo {author} {\bibfnamefont {N.~B.}\ \bibnamefont {Kovachki}},
  \bibinfo {author} {\bibfnamefont {K.}~\bibnamefont {Azizzadenesheli}},
  \bibinfo {author} {\bibfnamefont {B.}~\bibnamefont {Liu}}, \bibinfo {author}
  {\bibfnamefont {K.}~\bibnamefont {Bhattacharya}}, \bibinfo {author}
  {\bibfnamefont {A.}~\bibnamefont {Stuart}},\ and\ \bibinfo {author}
  {\bibfnamefont {A.}~\bibnamefont {Anandkumar}},\ }\bibfield  {title}
  {\enquote {\bibinfo {title} {Fourier neural operator for parametric partial
  differential equations},}\ }in\ \href
  {https://openreview.net/forum?id=c8P9NQVtmnO} {\emph {\bibinfo {booktitle}
  {International Conference on Learning Representations}}}\ (\bibinfo {year}
  {2021})\BibitemShut {NoStop}%
\bibitem [{\citenamefont {Wang}\ \emph {et~al.}(2025)\citenamefont {Wang},
  \citenamefont {Seidman}, \citenamefont {Sankaran}, \citenamefont {Wang},
  \citenamefont {Pappas},\ and\ \citenamefont
  {Perdikaris}}]{wang2025cvitcontinuousvisiontransformer}%
  \BibitemOpen
  \bibfield  {author} {\bibinfo {author} {\bibfnamefont {S.}~\bibnamefont
  {Wang}}, \bibinfo {author} {\bibfnamefont {J.~H.}\ \bibnamefont {Seidman}},
  \bibinfo {author} {\bibfnamefont {S.}~\bibnamefont {Sankaran}}, \bibinfo
  {author} {\bibfnamefont {H.}~\bibnamefont {Wang}}, \bibinfo {author}
  {\bibfnamefont {G.~J.}\ \bibnamefont {Pappas}},\ and\ \bibinfo {author}
  {\bibfnamefont {P.}~\bibnamefont {Perdikaris}},\ }\href
  {https://arxiv.org/abs/2405.13998} {\enquote {\bibinfo {title} {Cvit:
  Continuous vision transformer for operator learning},}\ } (\bibinfo {year}
  {2025}),\ \Eprint {https://arxiv.org/abs/2405.13998} {arXiv:2405.13998
  [cs.LG]} \BibitemShut {NoStop}%
\bibitem [{\citenamefont {Lu}\ \emph {et~al.}(2021)\citenamefont {Lu},
  \citenamefont {Jin}, \citenamefont {Pang}, \citenamefont {Zhang},\ and\
  \citenamefont {Karniadakis}}]{Lu_2021}%
  \BibitemOpen
  \bibfield  {author} {\bibinfo {author} {\bibfnamefont {L.}~\bibnamefont
  {Lu}}, \bibinfo {author} {\bibfnamefont {P.}~\bibnamefont {Jin}}, \bibinfo
  {author} {\bibfnamefont {G.}~\bibnamefont {Pang}}, \bibinfo {author}
  {\bibfnamefont {Z.}~\bibnamefont {Zhang}},\ and\ \bibinfo {author}
  {\bibfnamefont {G.~E.}\ \bibnamefont {Karniadakis}},\ }\bibfield  {title}
  {\enquote {\bibinfo {title} {Learning nonlinear operators via deeponet based
  on the universal approximation theorem of operators},}\ }\href
  {https://doi.org/10.1038/s42256-021-00302-5} {\bibfield  {journal} {\bibinfo
  {journal} {Nature Machine Intelligence}\ }\textbf {\bibinfo {volume} {3}},\
  \bibinfo {pages} {218–229} (\bibinfo {year} {2021})}\BibitemShut {NoStop}%
\bibitem [{\citenamefont {Pan}, \citenamefont {Brunton},\ and\ \citenamefont
  {Kutz}(2023)}]{JMLR:pan}%
  \BibitemOpen
  \bibfield  {author} {\bibinfo {author} {\bibfnamefont {S.}~\bibnamefont
  {Pan}}, \bibinfo {author} {\bibfnamefont {S.~L.}\ \bibnamefont {Brunton}},\
  and\ \bibinfo {author} {\bibfnamefont {J.~N.}\ \bibnamefont {Kutz}},\
  }\bibfield  {title} {\enquote {\bibinfo {title} {Neural implicit flow: a
  mesh-agnostic dimensionality reduction paradigm of spatio-temporal data},}\
  }\href {http://jmlr.org/papers/v24/22-0365.html} {\bibfield  {journal}
  {\bibinfo  {journal} {Journal of Machine Learning Research}\ }\textbf
  {\bibinfo {volume} {24}},\ \bibinfo {pages} {1--60} (\bibinfo {year}
  {2023})}\BibitemShut {NoStop}%
\bibitem [{\citenamefont {Raissi}, \citenamefont {Perdikaris},\ and\
  \citenamefont {Karniadakis}(2019)}]{RAISSI2019686}%
  \BibitemOpen
  \bibfield  {author} {\bibinfo {author} {\bibfnamefont {M.}~\bibnamefont
  {Raissi}}, \bibinfo {author} {\bibfnamefont {P.}~\bibnamefont {Perdikaris}},\
  and\ \bibinfo {author} {\bibfnamefont {G.}~\bibnamefont {Karniadakis}},\
  }\bibfield  {title} {\enquote {\bibinfo {title} {Physics-informed neural
  networks: A deep learning framework for solving forward and inverse problems
  involving nonlinear partial differential equations},}\ }\href
  {https://doi.org/https://doi.org/10.1016/j.jcp.2018.10.045} {\bibfield
  {journal} {\bibinfo  {journal} {Journal of Computational Physics}\ }\textbf
  {\bibinfo {volume} {378}},\ \bibinfo {pages} {686--707} (\bibinfo {year}
  {2019})}\BibitemShut {NoStop}%
\bibitem [{\citenamefont {Streckwall}, \citenamefont {Lindenau},\ and\
  \citenamefont {Bensch}(2007)}]{Streckwall07}%
  \BibitemOpen
  \bibfield  {author} {\bibinfo {author} {\bibfnamefont {H.}~\bibnamefont
  {Streckwall}}, \bibinfo {author} {\bibfnamefont {O.}~\bibnamefont
  {Lindenau}},\ and\ \bibinfo {author} {\bibfnamefont {L.}~\bibnamefont
  {Bensch}},\ }\bibfield  {title} {\enquote {\bibinfo {title} {Aircraft
  ditching: a free surface/free motion problem},}\ }\href
  {https://doi.org/https://doi.org/10.1016/S1644-9665(12)60025-9} {\bibfield
  {journal} {\bibinfo  {journal} {Archives of Civil and Mechanical
  Engineering}\ }\textbf {\bibinfo {volume} {7}},\ \bibinfo {pages} {177--190}
  (\bibinfo {year} {2007})}\BibitemShut {NoStop}%
\bibitem [{\citenamefont {{H. Schwarz, P. P. Lin, J.-P. M. Zemke and T.
  Rung}}(2025)}]{schwarz:2025}%
  \BibitemOpen
  \bibfield  {author} {\bibinfo {author} {\bibnamefont {{H. Schwarz, P. P. Lin,
  J.-P. M. Zemke and T. Rung}}},\ }\bibfield  {title} {\enquote {\bibinfo
  {title} {Disentangled latent spaces for reduced order models using
  deterministic autoencoders},}\ }\href
  {https://doi.org/https://doi.org/10.1016/j.compfluid.2025.106837} {\bibfield
  {journal} {\bibinfo  {journal} {Computers \& Fluids}\ }\textbf {\bibinfo
  {volume} {302}},\ \bibinfo {pages} {106837} (\bibinfo {year}
  {2025})}\BibitemShut {NoStop}%
\bibitem [{\citenamefont {Ha}, \citenamefont {Dai},\ and\ \citenamefont
  {Le}(2017)}]{ha2017hypernetworks}%
  \BibitemOpen
  \bibfield  {author} {\bibinfo {author} {\bibfnamefont {D.}~\bibnamefont
  {Ha}}, \bibinfo {author} {\bibfnamefont {A.~M.}\ \bibnamefont {Dai}},\ and\
  \bibinfo {author} {\bibfnamefont {Q.~V.}\ \bibnamefont {Le}},\ }\bibfield
  {title} {\enquote {\bibinfo {title} {Hypernetworks},}\ }in\ \href@noop {}
  {\emph {\bibinfo {booktitle} {International Conference on Learning
  Representations}}}\ (\bibinfo {year} {2017})\BibitemShut {NoStop}%
\bibitem [{\citenamefont {Sitzmann}, \citenamefont {Zollh\"{o}fer},\ and\
  \citenamefont {Wetzstein}(2019)}]{sitzmann:2019}%
  \BibitemOpen
  \bibfield  {author} {\bibinfo {author} {\bibfnamefont {V.}~\bibnamefont
  {Sitzmann}}, \bibinfo {author} {\bibfnamefont {M.}~\bibnamefont
  {Zollh\"{o}fer}},\ and\ \bibinfo {author} {\bibfnamefont {G.}~\bibnamefont
  {Wetzstein}},\ }\enquote {\bibinfo {title} {Scene representation networks:
  Continuous 3d-structure-aware neural scene representations},}\ in\ \href@noop
  {} {\emph {\bibinfo {booktitle} {Proceedings of the 33rd International
  Conference on Neural Information Processing Systems}}}\ (\bibinfo
  {publisher} {Curran Associates Inc.},\ \bibinfo {address} {Red Hook, NY,
  USA},\ \bibinfo {year} {2019})\BibitemShut {NoStop}%
\bibitem [{\citenamefont {Sitzmann}\ \emph {et~al.}(2020)\citenamefont
  {Sitzmann}, \citenamefont {Martel}, \citenamefont {Bergman}, \citenamefont
  {Lindell},\ and\ \citenamefont {Wetzstein}}]{sitzmann:2020}%
  \BibitemOpen
  \bibfield  {author} {\bibinfo {author} {\bibfnamefont {V.}~\bibnamefont
  {Sitzmann}}, \bibinfo {author} {\bibfnamefont {J.}~\bibnamefont {Martel}},
  \bibinfo {author} {\bibfnamefont {A.}~\bibnamefont {Bergman}}, \bibinfo
  {author} {\bibfnamefont {D.}~\bibnamefont {Lindell}},\ and\ \bibinfo {author}
  {\bibfnamefont {G.}~\bibnamefont {Wetzstein}},\ }\bibfield  {title} {\enquote
  {\bibinfo {title} {Implicit neural representations with periodic activation
  functions},}\ }in\ \href
  {https://proceedings.neurips.cc/paper_files/paper/2020/file/53c04118df112c13a8c34b38343b9c10-Paper.pdf}
  {\emph {\bibinfo {booktitle} {Advances in Neural Information Processing
  Systems}}},\ Vol.~\bibinfo {volume} {33},\ \bibinfo {editor} {edited by\
  \bibinfo {editor} {\bibfnamefont {H.}~\bibnamefont {Larochelle}}, \bibinfo
  {editor} {\bibfnamefont {M.}~\bibnamefont {Ranzato}}, \bibinfo {editor}
  {\bibfnamefont {R.}~\bibnamefont {Hadsell}}, \bibinfo {editor} {\bibfnamefont
  {M.}~\bibnamefont {Balcan}},\ and\ \bibinfo {editor} {\bibfnamefont
  {H.}~\bibnamefont {Lin}}}\ (\bibinfo  {publisher} {Curran Associates, Inc.},\
  \bibinfo {year} {2020})\ pp.\ \bibinfo {pages} {7462--7473}\BibitemShut
  {NoStop}%
\bibitem [{\citenamefont {Perez}\ \emph {et~al.}(2018)\citenamefont {Perez},
  \citenamefont {Strub}, \citenamefont {de~Vries}, \citenamefont {Dumoulin},\
  and\ \citenamefont
  {Courville}}]{Perez_Strub_deVries_Dumoulin_Courville_2018}%
  \BibitemOpen
  \bibfield  {author} {\bibinfo {author} {\bibfnamefont {E.}~\bibnamefont
  {Perez}}, \bibinfo {author} {\bibfnamefont {F.}~\bibnamefont {Strub}},
  \bibinfo {author} {\bibfnamefont {H.}~\bibnamefont {de~Vries}}, \bibinfo
  {author} {\bibfnamefont {V.}~\bibnamefont {Dumoulin}},\ and\ \bibinfo
  {author} {\bibfnamefont {A.}~\bibnamefont {Courville}},\ }\bibfield  {title}
  {\enquote {\bibinfo {title} {Film: Visual reasoning with a general
  conditioning layer},}\ }\href {https://doi.org/10.1609/aaai.v32i1.11671}
  {\bibfield  {journal} {\bibinfo  {journal} {Proceedings of the AAAI
  Conference on Artificial Intelligence}\ }\textbf {\bibinfo {volume} {32}}
  (\bibinfo {year} {2018}),\ 10.1609/aaai.v32i1.11671}\BibitemShut {NoStop}%
\bibitem [{\citenamefont {Dumoulin}\ \emph {et~al.}(2018)\citenamefont
  {Dumoulin}, \citenamefont {Perez}, \citenamefont {Schucher}, \citenamefont
  {Strub}, \citenamefont {Vries}, \citenamefont {Courville},\ and\
  \citenamefont {Bengio}}]{dumoulin2018feature-wise}%
  \BibitemOpen
  \bibfield  {author} {\bibinfo {author} {\bibfnamefont {V.}~\bibnamefont
  {Dumoulin}}, \bibinfo {author} {\bibfnamefont {E.}~\bibnamefont {Perez}},
  \bibinfo {author} {\bibfnamefont {N.}~\bibnamefont {Schucher}}, \bibinfo
  {author} {\bibfnamefont {F.}~\bibnamefont {Strub}}, \bibinfo {author}
  {\bibfnamefont {H.~d.}\ \bibnamefont {Vries}}, \bibinfo {author}
  {\bibfnamefont {A.}~\bibnamefont {Courville}},\ and\ \bibinfo {author}
  {\bibfnamefont {Y.}~\bibnamefont {Bengio}},\ }\bibfield  {title} {\enquote
  {\bibinfo {title} {Feature-wise transformations},}\ }\href
  {https://doi.org/10.23915/distill.00011} {\bibfield  {journal} {\bibinfo
  {journal} {Distill}\ } (\bibinfo {year} {2018}),\ 10.23915/distill.00011},\
  \bibinfo {note}
  {https://distill.pub/2018/feature-wise-transformations}\BibitemShut {NoStop}%
\bibitem [{\citenamefont {Rahaman}\ \emph {et~al.}(2019)\citenamefont
  {Rahaman}, \citenamefont {Baratin}, \citenamefont {Arpit}, \citenamefont
  {Draxler}, \citenamefont {Lin}, \citenamefont {Hamprecht}, \citenamefont
  {Bengio},\ and\ \citenamefont {Courville}}]{pmlr-v97-rahaman19a}%
  \BibitemOpen
  \bibfield  {author} {\bibinfo {author} {\bibfnamefont {N.}~\bibnamefont
  {Rahaman}}, \bibinfo {author} {\bibfnamefont {A.}~\bibnamefont {Baratin}},
  \bibinfo {author} {\bibfnamefont {D.}~\bibnamefont {Arpit}}, \bibinfo
  {author} {\bibfnamefont {F.}~\bibnamefont {Draxler}}, \bibinfo {author}
  {\bibfnamefont {M.}~\bibnamefont {Lin}}, \bibinfo {author} {\bibfnamefont
  {F.}~\bibnamefont {Hamprecht}}, \bibinfo {author} {\bibfnamefont
  {Y.}~\bibnamefont {Bengio}},\ and\ \bibinfo {author} {\bibfnamefont
  {A.}~\bibnamefont {Courville}},\ }\bibfield  {title} {\enquote {\bibinfo
  {title} {On the spectral bias of neural networks},}\ }in\ \href@noop {}
  {\emph {\bibinfo {booktitle} {Proceedings of the 36th International
  Conference on Machine Learning}}},\ \bibinfo {series} {Proceedings of Machine
  Learning Research}, Vol.~\bibinfo {volume} {97},\ \bibinfo {editor} {edited
  by\ \bibinfo {editor} {\bibfnamefont {K.}~\bibnamefont {Chaudhuri}}\ and\
  \bibinfo {editor} {\bibfnamefont {R.}~\bibnamefont {Salakhutdinov}}}\
  (\bibinfo  {publisher} {PMLR},\ \bibinfo {year} {2019})\ pp.\ \bibinfo
  {pages} {5301--5310}\BibitemShut {NoStop}%
\bibitem [{\citenamefont {Jacot}, \citenamefont {Gabriel},\ and\ \citenamefont
  {Hongler}(2018)}]{jacot:2018}%
  \BibitemOpen
  \bibfield  {author} {\bibinfo {author} {\bibfnamefont {A.}~\bibnamefont
  {Jacot}}, \bibinfo {author} {\bibfnamefont {F.}~\bibnamefont {Gabriel}},\
  and\ \bibinfo {author} {\bibfnamefont {C.}~\bibnamefont {Hongler}},\
  }\bibfield  {title} {\enquote {\bibinfo {title} {Neural tangent kernel:
  Convergence and generalization in neural networks},}\ }in\ \href
  {https://proceedings.neurips.cc/paper_files/paper/2018/file/5a4be1fa34e62bb8a6ec6b91d2462f5a-Paper.pdf}
  {\emph {\bibinfo {booktitle} {Advances in Neural Information Processing
  Systems}}},\ Vol.~\bibinfo {volume} {31},\ \bibinfo {editor} {edited by\
  \bibinfo {editor} {\bibfnamefont {S.}~\bibnamefont {Bengio}}, \bibinfo
  {editor} {\bibfnamefont {H.}~\bibnamefont {Wallach}}, \bibinfo {editor}
  {\bibfnamefont {H.}~\bibnamefont {Larochelle}}, \bibinfo {editor}
  {\bibfnamefont {K.}~\bibnamefont {Grauman}}, \bibinfo {editor} {\bibfnamefont
  {N.}~\bibnamefont {Cesa-Bianchi}},\ and\ \bibinfo {editor} {\bibfnamefont
  {R.}~\bibnamefont {Garnett}}}\ (\bibinfo  {publisher} {Curran Associates,
  Inc.},\ \bibinfo {year} {2018})\BibitemShut {NoStop}%
\bibitem [{\citenamefont {Tancik}\ \emph {et~al.}(2020)\citenamefont {Tancik},
  \citenamefont {Srinivasan}, \citenamefont {Mildenhall}, \citenamefont
  {Fridovich-Keil}, \citenamefont {Raghavan}, \citenamefont {Singhal},
  \citenamefont {Ramamoorthi}, \citenamefont {Barron},\ and\ \citenamefont
  {Ng}}]{tancik:2020}%
  \BibitemOpen
  \bibfield  {author} {\bibinfo {author} {\bibfnamefont {M.}~\bibnamefont
  {Tancik}}, \bibinfo {author} {\bibfnamefont {P.~P.}\ \bibnamefont
  {Srinivasan}}, \bibinfo {author} {\bibfnamefont {B.}~\bibnamefont
  {Mildenhall}}, \bibinfo {author} {\bibfnamefont {S.}~\bibnamefont
  {Fridovich-Keil}}, \bibinfo {author} {\bibfnamefont {N.}~\bibnamefont
  {Raghavan}}, \bibinfo {author} {\bibfnamefont {U.}~\bibnamefont {Singhal}},
  \bibinfo {author} {\bibfnamefont {R.}~\bibnamefont {Ramamoorthi}}, \bibinfo
  {author} {\bibfnamefont {J.~T.}\ \bibnamefont {Barron}},\ and\ \bibinfo
  {author} {\bibfnamefont {R.}~\bibnamefont {Ng}},\ }\bibfield  {title}
  {\enquote {\bibinfo {title} {Fourier features let networks learn high
  frequency functions in low dimensional domains},}\ }in\ \href@noop {} {\emph
  {\bibinfo {booktitle} {Proceedings of the 34th International Conference on
  Neural Information Processing Systems}}},\ \bibinfo {series and number}
  {NeurIPS '20}\ (\bibinfo  {publisher} {Curran Associates Inc.},\ \bibinfo
  {address} {Red Hook, NY, USA},\ \bibinfo {year} {2020})\BibitemShut {NoStop}%
\bibitem [{\citenamefont {Salimans}\ and\ \citenamefont
  {Kingma}(2016)}]{salimans:2016}%
  \BibitemOpen
  \bibfield  {author} {\bibinfo {author} {\bibfnamefont {T.}~\bibnamefont
  {Salimans}}\ and\ \bibinfo {author} {\bibfnamefont {D.~P.}\ \bibnamefont
  {Kingma}},\ }\bibfield  {title} {\enquote {\bibinfo {title} {Weight
  normalization: A simple reparameterization to accelerate training of deep
  neural networks},}\ }in\ \href
  {https://proceedings.neurips.cc/paper_files/paper/2016/file/ed265bc903a5a097f61d3ec064d96d2e-Paper.pdf}
  {\emph {\bibinfo {booktitle} {Advances in Neural Information Processing
  Systems}}},\ Vol.~\bibinfo {volume} {29},\ \bibinfo {editor} {edited by\
  \bibinfo {editor} {\bibfnamefont {D.}~\bibnamefont {Lee}}, \bibinfo {editor}
  {\bibfnamefont {M.}~\bibnamefont {Sugiyama}}, \bibinfo {editor}
  {\bibfnamefont {U.}~\bibnamefont {Luxburg}}, \bibinfo {editor} {\bibfnamefont
  {I.}~\bibnamefont {Guyon}},\ and\ \bibinfo {editor} {\bibfnamefont
  {R.}~\bibnamefont {Garnett}}}\ (\bibinfo  {publisher} {Curran Associates,
  Inc.},\ \bibinfo {year} {2016})\BibitemShut {NoStop}%
\bibitem [{\citenamefont {Kingma}\ and\ \citenamefont
  {Ba}(2015)}]{Kingma:2015}%
  \BibitemOpen
  \bibfield  {author} {\bibinfo {author} {\bibfnamefont {D.~P.}\ \bibnamefont
  {Kingma}}\ and\ \bibinfo {author} {\bibfnamefont {J.}~\bibnamefont {Ba}},\
  }\bibfield  {title} {\enquote {\bibinfo {title} {Adam: {A} method for
  stochastic optimization},}\ }in\ \href@noop {} {\emph {\bibinfo {booktitle}
  {Proceedings of the 3rd International Conference on Learning Representations
  (ICLR)}}}\ (\bibinfo {year} {2015})\BibitemShut {NoStop}%
\bibitem [{\citenamefont {Zintgraf}\ \emph {et~al.}(2019)\citenamefont
  {Zintgraf}, \citenamefont {Shiarli}, \citenamefont {Kurin}, \citenamefont
  {Hofmann},\ and\ \citenamefont {Whiteson}}]{pmlr-v97-zintgraf19a}%
  \BibitemOpen
  \bibfield  {author} {\bibinfo {author} {\bibfnamefont {L.}~\bibnamefont
  {Zintgraf}}, \bibinfo {author} {\bibfnamefont {K.}~\bibnamefont {Shiarli}},
  \bibinfo {author} {\bibfnamefont {V.}~\bibnamefont {Kurin}}, \bibinfo
  {author} {\bibfnamefont {K.}~\bibnamefont {Hofmann}},\ and\ \bibinfo {author}
  {\bibfnamefont {S.}~\bibnamefont {Whiteson}},\ }\bibfield  {title} {\enquote
  {\bibinfo {title} {Fast context adaptation via meta-learning},}\ }in\ \href
  {https://proceedings.mlr.press/v97/zintgraf19a.html} {\emph {\bibinfo
  {booktitle} {Proceedings of the 36th International Conference on Machine
  Learning}}},\ \bibinfo {series} {Proceedings of Machine Learning Research},
  Vol.~\bibinfo {volume} {97},\ \bibinfo {editor} {edited by\ \bibinfo {editor}
  {\bibfnamefont {K.}~\bibnamefont {Chaudhuri}}\ and\ \bibinfo {editor}
  {\bibfnamefont {R.}~\bibnamefont {Salakhutdinov}}}\ (\bibinfo  {publisher}
  {PMLR},\ \bibinfo {year} {2019})\ pp.\ \bibinfo {pages}
  {7693--7702}\BibitemShut {NoStop}%
\bibitem [{\citenamefont {Paszke}\ \emph {et~al.}(2019)\citenamefont {Paszke},
  \citenamefont {Gross}, \citenamefont {Massa}, \citenamefont {Lerer},
  \citenamefont {Bradbury}, \citenamefont {Chanan}, \citenamefont {Killeen},
  \citenamefont {Lin}, \citenamefont {Gimelshein}, \citenamefont {Antiga},
  \citenamefont {Desmaison}, \citenamefont {Kopf}, \citenamefont {Yang},
  \citenamefont {DeVito}, \citenamefont {Raison}, \citenamefont {Tejani},
  \citenamefont {Chilamkurthy}, \citenamefont {Steiner}, \citenamefont {Fang},
  \citenamefont {Bai},\ and\ \citenamefont {Chintala}}]{NEURIPS2019_9015}%
  \BibitemOpen
  \bibfield  {author} {\bibinfo {author} {\bibfnamefont {A.}~\bibnamefont
  {Paszke}}, \bibinfo {author} {\bibfnamefont {S.}~\bibnamefont {Gross}},
  \bibinfo {author} {\bibfnamefont {F.}~\bibnamefont {Massa}}, \bibinfo
  {author} {\bibfnamefont {A.}~\bibnamefont {Lerer}}, \bibinfo {author}
  {\bibfnamefont {J.}~\bibnamefont {Bradbury}}, \bibinfo {author}
  {\bibfnamefont {G.}~\bibnamefont {Chanan}}, \bibinfo {author} {\bibfnamefont
  {T.}~\bibnamefont {Killeen}}, \bibinfo {author} {\bibfnamefont
  {Z.}~\bibnamefont {Lin}}, \bibinfo {author} {\bibfnamefont {N.}~\bibnamefont
  {Gimelshein}}, \bibinfo {author} {\bibfnamefont {L.}~\bibnamefont {Antiga}},
  \bibinfo {author} {\bibfnamefont {A.}~\bibnamefont {Desmaison}}, \bibinfo
  {author} {\bibfnamefont {A.}~\bibnamefont {Kopf}}, \bibinfo {author}
  {\bibfnamefont {E.}~\bibnamefont {Yang}}, \bibinfo {author} {\bibfnamefont
  {Z.}~\bibnamefont {DeVito}}, \bibinfo {author} {\bibfnamefont
  {M.}~\bibnamefont {Raison}}, \bibinfo {author} {\bibfnamefont
  {A.}~\bibnamefont {Tejani}}, \bibinfo {author} {\bibfnamefont
  {S.}~\bibnamefont {Chilamkurthy}}, \bibinfo {author} {\bibfnamefont
  {B.}~\bibnamefont {Steiner}}, \bibinfo {author} {\bibfnamefont
  {L.}~\bibnamefont {Fang}}, \bibinfo {author} {\bibfnamefont {J.}~\bibnamefont
  {Bai}},\ and\ \bibinfo {author} {\bibfnamefont {S.}~\bibnamefont
  {Chintala}},\ }\bibfield  {title} {\enquote {\bibinfo {title} {Pytorch: An
  imperative style, high-performance deep learning library},}\ }in\ \href
  {http://papers.neurips.cc/paper/9015-pytorch-an-imperative-style-high-performance-deep-learning-library.pdf}
  {\emph {\bibinfo {booktitle} {Advances in Neural Information Processing
  Systems 32}}},\ \bibinfo {editor} {edited by\ \bibinfo {editor}
  {\bibfnamefont {H.~W.}\ \bibnamefont {et~al.}}}\ (\bibinfo  {publisher}
  {Curran Associates, Inc.},\ \bibinfo {year} {2019})\ pp.\ \bibinfo {pages}
  {8024--8035}\BibitemShut {NoStop}%
\bibitem [{\citenamefont {Bensch}, \citenamefont {Shigunov},\ and\
  \citenamefont {Söding}(2003)}]{bensch:2003}%
  \BibitemOpen
  \bibfield  {author} {\bibinfo {author} {\bibfnamefont {L.}~\bibnamefont
  {Bensch}}, \bibinfo {author} {\bibfnamefont {V.}~\bibnamefont {Shigunov}},\
  and\ \bibinfo {author} {\bibfnamefont {H.}~\bibnamefont {Söding}},\
  }\bibfield  {title} {\enquote {\bibinfo {title} {Computational method to
  simulate planned ditching of a transport airplane},}\ }in\ \href
  {https://doi.org/https://doi.org/10.1016/B978-008044046-0.50307-9} {\emph
  {\bibinfo {booktitle} {Computational Fluid and Solid Mechanics 2003}}},\
  \bibinfo {editor} {edited by\ \bibinfo {editor} {\bibfnamefont
  {K.}~\bibnamefont {Bathe}}}\ (\bibinfo  {publisher} {Elsevier Science Ltd},\
  \bibinfo {address} {Oxford},\ \bibinfo {year} {2003})\ pp.\ \bibinfo {pages}
  {1251--1254}\BibitemShut {NoStop}%
\end{thebibliography}%

\end{document}